\documentclass[11pt]{article}

\def\tr{\;{\rm tr}\;}

\def\bra{\langle}   \def\ket{\rangle}
\def\pr{\prime}

\newcommand{\tl}[1]{\tilde{#1}}

\newcommand{\dd}[2]{\frac {\partial #1}{\partial #2}}
\newcommand{\pdr}{\partial}

\newcommand{\beq}{\begin{eqnarray}}
\newcommand{\eeq}{\end{eqnarray}}

\newcommand{\half}{\frac{1}{2}}

\newcommand{\ov}[1]{\frac{1}{#1}}

\def\a{\alpha}         \def\g{\gamma}       \def\G{\Gamma}
\def\d{\delta}      \def\D{\Delta}  \def\eps{\epsilon} 
         \def\k{\kappa} \def\la{\lambda}      \def\La{\Lambda}

     \def\s{\sigma}

\newcommand{\N}{{1 \over N}}

\newcommand{\Ntr}{{{\rm tr} \over N}}

\newcommand{\bfD}{{\bf D}}
\newcommand{\bfS}{{\bf S}}

\begin{document}

\begin{titlepage}

\title{\normalsize \hfill ITP-UU-06/20  \\ \hfill SPIN-06/17
\\ \hfill {\tt hep-th/0606224}\\ \vskip 0mm \Large\bf
Multi-matrix loop equations: algebraic \& differential structures
and an approximation based on deformation quantization}

\author{Govind S. Krishnaswami}
\date{\normalsize Institute for Theoretical Physics \& Spinoza Institute \\
Utrecht University, Postbus 80.195, 3508 TD, Utrecht, The
Netherlands
\smallskip \\ e-mail: \tt g.s.krishnaswami@phys.uu.nl \\ June 22, 2006}

\maketitle

\begin{quotation} \noindent {\large\bf Abstract } \medskip \\

Large-$N$ multi-matrix loop equations are formulated as quadratic
difference equations in concatenation of gluon correlations. Though
non-linear, they involve highest rank correlations linearly. They
are underdetermined in many cases. Additional linear equations for
gluon correlations, associated to symmetries of action and measure
are found. Loop equations aren't differential equations as they
involve left annihilation, which doesn't satisfy the Leibnitz rule
with concatenation. But left annihilation is a derivation of the
commutative shuffle product. Moreover shuffle and concatenation
combine to define a bialgebra. Motivated by deformation
quantization, we expand concatenation around shuffle in powers of
$q$, whose physical value is $1$. At zeroth order the loop equations
become quadratic PDEs in the shuffle algebra. If the variation of
the action is linear in iterated commutators of left annihilations,
these quadratic PDEs linearize by passage to shuffle reciprocal of
correlations. Remarkably, this is true for regularized versions of
the Yang-Mills, Chern-Simons and Gaussian actions. But the linear
equations are underdetermined just as the loop equations were. For
any particular solution, the shuffle reciprocal is explicitly
inverted to get the zeroth order gluon correlations. To go beyond
zeroth order, we find a Poisson bracket on the shuffle algebra and
associative $q$-products interpolating between shuffle and
concatenation. This method, and a complementary one of deforming
annihilation rather than product are shown to give over and
underestimates for correlations of a gaussian matrix model.

\end{quotation}

\vfill \flushleft




\thispagestyle{empty}

\end{titlepage}

\eject

{\scriptsize \tableofcontents}

\section{Introduction}
\label{s-intro}

\subsection{General Remarks}
\label{s-gen-rmks}

Approximation methods in physics are often usefully organized as an
expansion in a dimensionless parameter. As is well known, at first
sight, quantum Yang-Mills theory does not have any such expansion
parameter since the dimensionless coupling $g^2$ of the classical
theory is determined in terms of the ratio $Q^2 \over \La^2$ where
$Q^2$ is the momentum transferred to a hadronic system by an
external (say electroweak) current. $\La$ (say $\La_{\rm QCD}$) is
the dimensional parameter arising via dimensional transmutation and
renormalization. The success of an expansion in inverse
(logarithmic) powers of $Q^2 \over \La^2$ is, however, crucially
dependent on the asymptotic freedom of the theory for large values
of this parameter\cite{politzer}. Thus, this expansion (perturbative
QCD), which is the analogue of the Born approximation of atomic
physics, though spectacularly successful at high momentum transfers,
is not particularly useful to describe `intrinsic' properties of
hadrons in the absence of an external probe transferring a large
momentum \cite{witten-N-atomic-particle}.

What about $\hbar$ as an expansion parameter for quantum Yang-Mills
theory around its classical limit? This is a bad starting point,
since all variables, not just gauge-invariant ones, stop fluctuating
in this limit. Since $\hbar$ can be absorbed into $g^2$, the `loop'
expansion in powers of $\hbar$ around the trivial solution to
classical Yang-Mills theory is the same as perturbative QCD. Thus,
it is useful only at high momentum transfers.

As observed by 't Hooft\cite{thooft-large-N}, $1/N$ of the gauge
group $SU(N)$ is an expansion parameter for quantum Yang-Mills
theory, holding $\la = g^2 N$ fixed. There are many
indications\cite{witten-baryons-N} that $N \to \infty$ is a good
approximation to the quantum theory. Moreover, it is a classical
limit where fluctuations in gauge-invariant variables alone vanish.
Despite effort, the $1/N$ expansion has not been as quantitatively
successful as perturbative QCD was in the high energy regime. The
success of the loop expansion lay in the availability of explicit
solutions to classical Yang-Mills theory around which to expand (eg.
flat connections, Euclidean instantons). By contrast, we don't know
the zeroth order solution of large $N$ Yang-Mills theory around
which to perform a $1/N$ expansion. Difficulties are encountered in
each of the many ways of formulating the large $N$ limit of
Yang-Mills theory: summing an infinite class of planar
diagrams\cite{thooft-large-N}, solving the Makeenko-Migdal equations
for Wilson
loops\cite{makeenko-migdal-eqn,migdal-phys-rpts,makeenko-book} or
solving the factorized Schwinger-Dyson equations for gluon
correlations. It would really help to have yet another dimensionless
expansion parameter, to organize an approximate solution of $N =
\infty$ Yang-Mills theory.

The strategy of looking for an expansion parameter over and above
$1/N$ has found success in maximally super-symmetric Yang-Mills
theory. In some sectors of the ${\cal N}=4$ theory, an expansion
around small values of the ratio of 't Hooft coupling to square of
$R$-charge ($\la \over J^2$) has been developed\cite{bmn}. An analog
of this for the non-supersymmetric theory would be useful. But since
there is no such obvious expansion parameter, we will invent one
based on deeper mathematical structures of the theory.

Inspiration for a possible approximation comes from atomic physics,
as emphasized by Rajeev\cite{rajeev-neo-classical}. The Hartree-Fock
approximation for many-electron atoms is analogous to the $N \to
\infty$ limit of Yang-Mills theory, since it can be formulated as
the limit in which the number of replicas of each electron ($N$)
tends to infinity \cite{rajeev-qhd}. In general, the Hartree-Fock
equations are difficult to solve since they involve the electron
density matrix, which is a projection operator. However, after $N
\to \infty$ it is possible to take a semiclassical limit based on
deformation quantization. These limits do not commute. At zeroth
order this leads to the Thomas-Fermi non-linear ODE whose solution
gives a good first approximation to the charge density of a
many-electron atom \cite{rajeev-neo-classical}. Can something
similar work for large $N$ Yang-Mills theory?

The approximation method studied in this paper is based on the
observation that even in the `classical' large-$N$ limit, the
equations of matrix models and Yang-Mills theory still involve
non-commutative concatenation products. It should be possible to
take a further `classical' limit, where they are approximated by
commutative products by analogy with deformation quantization. In
our case, the parameter controlling this further classical limit is
a deformation parameter whose physical value is $q=1$.

Another lesson from the formulation of Hartree-Fock theory as the
limit of a large number of electron replicas, is that the physical
value of an expansion parameter need not be small for the expansion
to be practically successful. Indeed, the physical number of
replicas of the electron is $N = 1$ and yet, Hartree-Fock, which
corresponds to $N=\infty$, provides a good first approximation as
part of a $1/N$ expansion! Another example is the $\d$ expansion of
Bender and collaborators\cite{bender-delta-QED}. Applied to QED, it
can be regarded as an expansion in the number of identically charged
electron species whose physical value is $\d=1$. Yet an expansion in
powers of $\d$ is accurate. It has also been successfully applied to
a variety of other non-linear equations.

Another possible expansion parameter is the inverse number of
space-time dimensions $1/d$. However, we do not yet know of any
useful formulation of the $d \to \infty$ limit of large $N$
Yang-Mills theory that is a simplification. This is again motivated
by atomic physics, where the $d \to \infty$ limit in the zero
angular momentum sector is a non-relativistic $O(d)$ vector model
for position vectors of electrons. This provides a spectacularly
good approximation to the binding energies of many-electron atoms in
a $1/d$ expansion, as shown by Herschbach and
collaborators\cite{herschbach-2-elec}.

\subsection{Loop Equations of Large-$N$ Matrix Models}
\label{s-large-N-mat-mod-loop-eq}

A primary aim in the study of a Euclidean large-$N$ multi-matrix
model is to determine its factorized correlations. They satisfy
quantum corrected equations of motion, which are factorized
Schwinger-Dyson or loop equations (LE). We formulate these in a way
that makes manifest some algebraic and differential structures they
share with the Makeenko-Migdal equations of $N=\infty$ Yang-Mills
theory\cite{makeenko-migdal-eqn,migdal-phys-rpts,makeenko-book}. In
particular, they are not differential equations, due to a mismatch
between the differential and product structures. Though infinite in
number and quadratically non-linear, we show that they have a
hierarchical structure whereby the highest rank correlations in any
equation only appear linearly. However, we show they are
underdetermined in many interesting cases. We identify additional
equations which a naive passage to the large $N$ limit misses. They
are conditions implied by invariance of matrix integrals for
correlations, under transformations leaving both action and measure
invariant, possibly up to $1/N^2$ corrections (eg. BRST
transformations). However, the additional equations are not
implemented, so the underdeterminacy of the loop equations is not
satisfactorily resolved. On the other hand, we exploit the algebraic
and differential structures to propose an approximation scheme for a
class of $\La$-(multi)-matrix models motivated by the Lagrangian of
Yang-Mills theory,
    \beq
    {\cal L} &=& \tr\bigg\{\half \pdr_\mu A_\nu(\pdr^\mu A^\nu ~-~ \pdr^\nu A^\mu)
    ~-~ ig \pdr_\mu A_\nu [A^\mu,A^\nu] ~-~ {g^2 \over 4} [A_\mu,A_\nu][A^\mu,A^\nu]
    \cr && ~+~ \ov{2 \xi} (\pdr^\mu A_\mu)^2 ~+~ \pdr_\mu \bar c ~\pdr^\mu c
    ~-~  ig \pdr_\mu \bar c ~[A^\mu,c]  \bigg\}.
    \label{e-gauge-fixed-ym-action}
    \eeq
The primary virtue of the scheme is that at zeroth order, it turns
the non-linear loop equations into linear PDEs. Prominent in this
class of models are those whose action is a linear sum of
    \beq
    S_{G} = \half \tr C^{ij} A_i A_j,~~
    S_{CS} = {2 i \k \over 3} \tr C^{ijk} A_i [A_j,A_k] ~~\&~
    S_{YM} = -\ov{4\a} \tr [A_i, A_j][A_k,A_l] g^{ik} g^{jl}.
    \label{e-gauss-cs-ym-actions}
    \eeq
In the first two cases, we allow $A_i$ to denote either
gluon(hermitian complex) or ghost(grassmann) matrices\footnote{We
assume there are an equal number of ghost and anti-ghost matrices in
each term, as in Yang-Mills theory.}. Though they arise from terms
with $2,1$ and $0$ derivatives in the Yang-Mills action, these
matrix models may be called Gaussian, Chern-Simons and Yang-Mills
models since they also include the zero momentum limits of the
corresponding field theories. The indices $i,j,k,l$ are short for
position and polarization quantum numbers, while color indices are
suppressed. It may be possible to fruitfully think of Yang-Mills
theory as a grand limiting case of such matrix models for
appropriate integral kernels $C^{ij}, C^{ijk}$ and $g^{ij}$ when the
indices become continuous. Matrix models and field theories of this
type also arise in dimensional reductions of Yang-Mills theory to
$2$ or fewer space-time dimensions. Here we consider bosonic matrix
models, the extension of our results to models with ghost matrices
will be treated in \cite{gsk-ghost-shuffle-concat}.

{\flushleft \bf Summary of results and organization:} In section
\ref{s-factorized-loop-eqns} we obtain the large-$N$ loop
equations\footnote{Small letters $i$ denote single indices, capitals
denote multi-indices $I = i_1 i_2 \cdots i_n$ and $|I|$ denotes the
number of indices in a multi-index. Repeated upper and lower indices
are summed. $\d^I_J$ is $1$ if $I =J$ and zero otherwise.} $|iJ|
S^{Ji} G_{JI} = \d^{I_1 i I_2}_I G_{I_1} G_{I_2}$ for gluon
correlations $G_I = \bra \N \tr A_I \ket$ of a hermitian
multi-matrix model with action $\tr S(A) = \tr S^I A_I$. In section
\ref{s-non-uniq-of-loop-eq-solns} we show that the loop equations
are underdetermined in some interesting cases, though they determine
infinitely many higher rank correlation in terms of lower rank
correlations. In section \ref{s-additional-eqns} we obtain
additional equations associated with symmetries of both measure and
action, which are easily overlooked in passing to the large-$N$
limit. In section \ref{s-loop-eqns-left-ann-conc} the loop equations
are reformulated in terms of the series $G(\xi) = G_I \xi^I$, where
$\xi^i$ are non-commuting sources:
    \beq
    \sum_{n \geq 0} (n+1) S^{j_1 \cdots j_n i} D_{j_n} \cdots D_{j_1} G(\xi) =  G(\xi)
        \xi^i G(\xi)  {\rm ~~~ or ~~~} {\cal S}^i G(\xi ) = G(\xi)
        \xi^i G(\xi).
    \eeq
The linear term (variation of action) is written in terms of left
annihilation operators $D_i$. The quadratic term in gluon
correlations involves the concatenation product. It is the variation
of the matrix model measure and is universal, independent of the
action. However, left annihilation does not satisfy the Leibnitz
rule with respect to concatenation, and to make things worse,
concatenation is non-commutative. Due to this mismatch, the loop
equations are not differential equations in the ordinary sense. On
the other hand, there is another natural product between gluon
correlations, the shuffle product (section
\ref{s-sh-prod-from-wilson-loops}), which arises from the
expectation value of point-wise products of Wilson loops. It turns
out that left annihilation is a derivation of the shuffle product.
Moreover, there is a democratic version of left annihilation, full
annihilation, that is a derivation of concatenation (section
\ref{s-derivations-of-sh-conc}). Furthermore, concatenation and
shuffle combine to form a bialgebra (appendices
\ref{a-conc-sh-and-coprods} and \ref{a-bialgebra-structure}).

These algebraic and differential structures along with ideas from
deformation quantization suggest a possible approximation scheme for
the loop equations. The idea is to remedy the above mismatch by
expanding the non-commutative concatenation product in a series
around the commutative shuffle product so that at zeroth order,
concatenation is replaced by shuffle and the loop equations become
quadratically non-linear inhomogeneous PDEs in an infinite
dimensional space spanned by words in $\La$ letters. Thus, the
approximation scheme involves the introduction of a deformation
parameter controlling the amount by which the loop equations for
gluon and ghost correlations fail to be partial differential
equations. The physical value of our dimensionless expansion
parameter $q$ is $1$.

A further remarkable simplification occurs in models whose action is
such that ${\cal S}^i$ is a derivation of the shuffle product. These
are models in which ${\cal S}^i$ is a linear combination of iterated
commutators of $D_i$ and include the zero-momentum Gaussian,
Chern-Simons and Yang-Mills models as well as their field theoretic
counterparts as examples (section \ref{s-derivation-property}). In
these cases, the passage from $G(\xi)$ to its shuffle-reciprocal
$F(\xi) = F_I \xi^I$ turns the non-linear PDEs into a system of
linear equations for the $F_I$ (section
\ref{s-multi-mat-zeroth-ord-q}). We obtain an explicit formula for
$G_I$ in terms of $F_J$ so that once the linear equations are
solved, the ${\cal O}(q^0)$ gluon correlations can be obtained. This
is illustrated for the zero-momentum Gaussian (section
\ref{s-gauss-multi-mat-zeroth-order-in-q}), Chern-Simons (section
\ref{s-cs-zeroth-ord-q}) and Yang-Mills (section
\ref{s-ym-zeroth-ord-q}) multi-matrix models. For the Gaussian, the
linear equations have a unique solution which provides a first
approximation to the exact large $N$ correlations. But for the other
examples, the equations are underdetermined just as the original
loop equations were and we exhibit infinite classes of solutions. It
remains to find and implemented the additional constraints on
correlations, such as those associated to symmetries of action and
measure.

In section \ref{s-multi-mat-beyond-zeroth-order} we take the first
steps to extend the approximation scheme beyond zeroth order. This
requires us to find an expansion for concatenation around the
shuffle product. Such a formula would be loosely analogous to the
associative $*$-product expressions of deformation quantization. We
obtain two partial results in this direction. First, we find a one
parameter family of associative $q$-products that interpolates
between commutative shuffle $(q=0)$ and non-commutative
concatenation $(q=1)$. Moreover, by taking $q$ to be infinitesimal,
we obtain a Poisson bracket on the shuffle algebra.

In sections \ref{s-multi-mat-zeroth-ord-p} and
\ref{s-q-annihilation} we briefly investigate another approximation
scheme for the loop equations that involves expanding the left
annihilation around full annihilation, holding the concatenation
product fixed. Though similar in spirit to the main approximation
scheme of the paper, it has the potential to give a complementary
estimate for correlations as shown by its application to $1$-matrix
models.

Section \ref{s-approx-meth-one-mat}, is devoted to $1$-matrix
models. In this case, both concatenation and shuffle are
commutative, and an explicit `star product' formula is obtained for
the expansion of the former around the latter (section
\ref{s-1-mat-q-deformed-prod}). In section
\ref{s-1-mat-q-deformed-annihilation} an expansion for the left
annihilation as a series in powers of full annihilation is obtained.
These lead to two different approximation methods for the $1$-matrix
loop equations, involving either a deformation of the product or the
annihilation operator. Both schemes are applied to the Gaussian
(section \ref{s-1-mat-gaussian}), which is the only $1$-matrix model
for which ${\cal S}^i$ has the derivation property. While deforming
the product overestimates correlations, deforming the annihilation
operator underestimates them.

{\flushleft \bf Background on Literature:} There are several
complementary approaches to the loop equations of matrix models.
First, they are formulated in different ways: resolvents of
matrices, gluon correlations, planar diagrams, Wilson loops etc.
Different approaches to multi-matrix models can be broadly
categorized by the mathematical structures that play a significant
role. A major portion of the literature (eg.
\cite{staudacher-2-mat,kazakov-marshakov,eynard-loop-eqn-chain,chekov-eynard-orantin-2-mat})
is devoted to exact solutions for certain observables of specific
(e.g. $1$-, $2$- and chain-type) matrix models, their multi-cut
solutions and summing their $1/N$ expansion. This involves
connections to integrable systems, algebraic geometry and conformal
field theory. Another approach exploits the connections to
non-commutative probability theory (eg.
\cite{douglas-li,gopakumar-gross,entropy-var-ppl,hamiltonian-mat-mod-fisher-info}).
Yet another point of view seeks to exploit a hidden BRST symmetry
\cite{alfaro}. A cohomological interpretation of the loop equations
and a variational principle for them was presented in
\cite{entropy-var-ppl}. The viewpoint in this paper is distinguished
by its use of algebraic and differential structures and connections
to deformation quantization. Its physics roots lie in the early work
of Makeenko and Migdal\cite{makeenko-migdal-eqn,migdal-phys-rpts},
Cvitanovic et. al.\cite{Cvitanovic:1980jz,Cvitanovic:1982rf}, loop
space formalism for gauge theories
\cite{mandelstam-qed-loops,gambini-trias,tavares,gambini-pullin-book},
and the more recent investigations of Rajeev and coworkers
\cite{rajeev-turgut-der-free-alg,lee-rajeev-closed,coh-interpret-mm-eqn,entropy-var-ppl,hamiltonian-mat-mod-fisher-info,rajeev-neo-classical}.
Some structures used in our constructions (eg. shuffle products and
their deformations) appear in the mathematics literature on calculus
of loop space due to Chen \cite{chen-1}, the theory of free Lie
algebras \cite{free-lie-alg} and the deformation theory of (Hopf)
algebras \cite{q-shuffle,rosso-quantum-sh}. A feature of the present
work is that we do not make any a priori restriction to a subclass
of correlations (eg. `mixed' or `unmixed') as is often assumed in
the literature.

\section{Algebraic structure of loop equations of multi-matrix models}
\label{s-alg-str-loop-eq}

\subsection{Factorized loop equations for gluon correlation tensors}
\label{s-factorized-loop-eqns}

We begin by obtaining the loop equations of a bosonic multi-matrix
model in terms of gluon correlation tensors. This is convenient to
study their algebraic structures and permits treatment of all
factorized $N=\infty$ correlations without restriction. Consider a
Euclidean $\La$-matrix model with polynomial action $\tr S(A) = \tr
S^J A_J$. Let $\Phi_I = \N \tr A_I$ denote the `loop' variable. The
partition function and gluon correlations are
 \beq
    Z = \int \Pi_j dA_j  e^{-N \tr S(A)} {\rm ~~and~~}
    \bra \Phi_{K_1} \cdots \Phi_{K_n} \ket = \ov{Z} \int \Pi_j dA_j  e^{-N \tr S(A)}
    \Phi_{K_1} \cdots \Phi_{K_n}.
    \label{e-def-of-partn-fn-correlations}
 \eeq
$G_K  = \lim_{N \to \infty} \bra \Phi_K \ket$ are the gluon
correlations of interest in the large-$N$ limit. Here $A_i =
A_i^\dag,~ 1 \leq i \leq \La$ are $N \times N$ hermitian matrices.
The tensors $S^I$ are the `coupling tensors' defining the theory.
Due to the trace, the only part of $S^I$ that contributes is its
cyclic projection, so assume that $S^I$ are cyclically symmetric,
$S^{Ii} = S^{iI}$ for all $i, I$. Gluon correlation tensors $G_I$
are also cyclically symmetric. Additionally, assume $S^I$ are chosen
such that $(S^I)^* = S^{\bar I}$ where ${\bar I}$ is the word with
indices reversed\footnote{This is satisfied by examples such as the
Gaussian, Yang-Mills and Chern-Simons theories, see Sec
\ref{s-derivation-property}.}. This, along with hermiticity of $A_i$
ensures that $\tr S(A)$ is real. In turn, this implies that $G_I^* =
G_{\bar I}$. To see this, recall that for any complex matrix $M$,
$(\tr M)^* = \tr M^\dag$ and apply this to $M = A_I$ and use
hermiticity of $A_i$. For the Gaussian, all $S^I =0$ except $S^{ij}$
which may be taken as a (positive) real-symmetric matrix.

The Schwinger-Dyson equations(SDE) are constraints on $\bra
\Phi_{K_1} \cdots \Phi_{K_n} \ket$ implied by invariance of the
matrix integral under an infinitesimal (but non-linear) change of
integration variable
    \beq
    [A_i]^b_a \mapsto [A_i^\pr]^b_a = [A_i]^b_a + v^I_i
    [A_I]^b_a, {\rm ~~ where~~ } v^I_i
        {\rm ~~are~infinitesimal ~real~ parameters}.
    \eeq
Under this change of variable, the infinitesimal changes in
$\Phi_K$, the action and the measure are
    \beq
    \Phi_K &\mapsto& \Phi_K + \d_K^{L i M} v_i^I
        \Phi_{L I M}, \cr
    e^{-N \tr S^J A_J} &\mapsto& e^{-N \tr S^J A_J}
        (1-N^2 v_i^I S^{J_1 i J_2} \Phi_{J_1 I J_2} ),
        \cr
    \det\bigg( \dd{[A^\pr_i]^a_b}{[A_j]^c_d} \bigg) &=&
        1 + N^2 v_i^I \d_I^{I_1 i I_2} \Phi_{I_1} \Phi_{I_2}.
    \eeq
Invariance of $\bra \Phi_{K_1} \cdots \Phi_{K_n} \ket$ to linear
order in $v^I_i$ implies the SDE\footnote{Sometimes called Virasoro
constraints in string models or Ward identities. Ward identities
seems more appropriate to the special case where the change of
integration variable was a gauge or BRST transformation.}
    \beq
    v^I_i S^{J_1 i J_2} \bra \Phi_{J_1 I J_2} \ket = v^I_i  \d_I^{I_1 i
    I_2} \bra \Phi_{I_1} \Phi_{I_2} \ket + {v^I_i \over N^2} \sum_{p=1}^n \d_{K_p}^{L_p i M_p} \bra
    \Phi_{L_p I M_p} \ket, ~~~~ \forall~~ K_p {\rm ~~and~~} n=0,1,2,\ldots
    \label{e-finite-N-SD-eq}
    \eeq
So far we have not made any approximation. In the large $N$ limit,
expectation values of $U(N)$ invariants factorize $\bra \Phi_{I_1}
\Phi_{I_2} \ket = \bra \Phi_{I_1} \ket \bra \Phi_{I_2} \ket$
\cite{makeenko-book}. Naively, the leading factorized
Schwinger-Dyson or loop equations (LE), which are a closed system
for $G_I$, are
    \beq
    v^I_i S^{J_1 i J_2} G_{J_1 I J_2} = v^I_i \delta^{I_1 i I_2}_I G_{I_1}
    G_{I_2} ~~~ \forall ~~ v
    \label{e-loop-eqns-for-general-vector-field}
    \eeq
These infinitesimal changes of variable are associated to vector
fields $L_v = v^I_i L^i_I$ whose action on $G_J$ is given by $L^i_I
G_J = \d_J^{J_1 i J_2} G_{J_1 I J_2}$. In particular, choosing the
components of the vector fields $v^I_i$ to be non-vanishing only for
a single $(i,I)$, we get the loop equations
    \beq
    S^{J_1 i J_2} G_{J_1 I J_2} = \delta^{I_1 i I_2}_I G_{I_1}
    G_{I_2} ~~~ \forall ~~ I,~~i.
    \eeq
Using cyclicity of $S^I$ and $G_I$ we get
    \beq
    |iJ|~ S^{Ji} G_{JI} = \delta^{I_1 i I_2}_I G_{I_1} G_{I_2}
    ~~\forall~~ I, ~i.
    \label{e-loop-eqns}
    \eeq
LE (\ref{e-loop-eqns}) relate a changes in (expectation values of)
action and measure under the action of $L^i_I$. However, there may
be vector fields $L_v$ (i.e. choices of $v^I_i$) for which both
sides of \ref{e-loop-eqns-for-general-vector-field}
vanish\footnote{Note that this may happen even if there is no
$(i,I)$ for which both sides of \ref{e-loop-eqns} vanish.}. In that
case, the leading equation in the large $N$ limit is different from
\ref{e-loop-eqns} (see section \ref{s-additional-eqns}).

We seek solutions to \ref{e-loop-eqns} among cyclic symmetric
tensors $G_I$ satisfying $G_I^* = G_{\bar I}$ and $G_\emptyset
\equiv G_0 =1$, where $\emptyset$ is the empty string. Note that the
LE may make sense even when the matrix integrals don't seem to
converge, as for a cubic action. When analogues of
(\ref{e-loop-eqns}) are formulated for Wilson loops in a gauge
theory\cite{makeenko-migdal-eqn}, they are called Makeenko-Migdal
equations (notice the resemblance between (\ref{e-loop-eqns}) and
(\ref{e-makeenko-migdal-eqns}))
    \beq
    \d^x_\mu {\d \over \d \s_{\mu \nu}(x) } W(C) = \la \oint_C
    dy_\nu \d^{(4)}(x-y) W(C_{yx}) W(C_{xy}).
    \label{e-makeenko-migdal-eqns}
    \eeq

\subsection{Underdetermined nature of loop equations and examples}
\label{s-non-uniq-of-loop-eq-solns}

Given an action $S(A)$, $G_I$ are uniquely defined by
(\ref{e-def-of-partn-fn-correlations}) provided the integrals
converge. As examples below show, the large-$N$ LE
(\ref{e-loop-eqns}) do not determine $G_I$ uniquely in general. In
section \ref{s-additional-eqns} we obtain additional large-$N$ SDE
involving $G_I$ that were not accounted for in the passage from
(\ref{e-finite-N-SD-eq}) to (\ref{e-loop-eqns}). But even these may
not be sufficient to fix the $G_I$.

Consider first $\La = 1$ matrix models whose LE are got by
restricting (\ref{e-loop-eqns}) to a single matrix. Suppose $\tr
S(A) = \tr \sum_{l=1}^m S_l A^l$ is an $m^{\rm th}$ order
polynomial, then if $G_k = \bra \Ntr A^k \ket$
    \beq
    \sum_{l=1}^m l S_l G_{k+l} = \sum_{r,s \geq 0,~ r+s = k} G_r G_s,
    ~~~ for ~~ k=-1,0,1,\cdots.
    \label{e-1-mat-loop-eq}
    \eeq
The LE listed sequentially are
    \beq
    k=-1: && S_1 + 2 S_2 G_1  + \cdots + m S_m G_{m-1} =0,
        \cr
    k=0: && S_1 G_1 + 2 S_2 G_2 + \cdots + m S_m G_{m} = 1,
        \cr
    k=1: && S_1 G_2 + 2 S_2 G_3 + \cdots + m S_m G_{m+1} = 2G_1,
        \cr
    k=2: && S_1 G_3 + 2 S_2 G4 + \cdots + m S_m G_{m+2} = 2 G_2 +
        G_1^2,
    ~~~~ \ldots
    \eeq
We see that in the $k^{\rm th}$ equation, the highest rank
correlation $G_{m+k}$ appears linearly ($S_m \ne 0$) and may be
determined in terms of lower rank correlations. For a Gaussian
$(m=2)$ (\ref{e-1-mat-loop-eq}) determine all moments. More
generally, the LE determine higher moments $G_{m-1},G_m, G_{m+1},
\ldots$ in terms of $m-2$ undetermined lower moments $G_1, \ldots
G_{m-2}$. However, among $G_1, \cdots G_{m-2}$, the odd ones must
vanish if the action is even. Observe that this is associated with
the $[A]^a_b \mapsto -[A]^a_b$ symmetry of an even action and of the
measure if $N \to \infty$ through even values. Such transformations
provide additional equations missed out by the LE.

For multi-matrix models, suppose $S(A)$ is an $m^{\rm th}$ order
polynomial, i.e $S^J = 0$ if $|J| > m$ and $\exists ~J$ with $|J|=m$
such that $S^J \ne 0$. Then the loop equation $|iJ| S^{Ji} G_{JI} =
\d^{I_1 i I_2}_I G_{I_1} G_{I_2}$ for any fixed $I$ and $i$ involves
correlations with highest rank ($|I|+m-1$) only linearly. Of course,
there are several correlations with a given rank and several
equations for fixed $|I|$. If all $G_K$ up to $|K| \leq r$ are
known, we have a system of inhomogeneous linear equations for
correlations of rank $r+1$. For the Gaussian $\tr S(A) = \half \tr
C^{ij} A_i A_j$, these are just recursion relations $G_{iI} = C_{ij}
\d^{I_1 j I_2}_I G_{I_1} G_{I_2}$ where $C_{ij} C^{jk} = \d^k_i$.
Their unique solution for all correlations is given by the planar
version of Wick's theorem, which is a sum over all non-crossing
partitions of $iI$ into pairs. But for many interesting cubic and
higher order actions, the LE are underdetermined even by comparison
with $1$-matrix models. Not only are $G_K$ for $|K| \leq m-2$ left
undetermined, many higher rank correlations are also not determined
in terms of them. Consider two examples: a quartic $2$-matrix model
and the Chern-Simons $3$-matrix model.

\subsubsection{Quartic $2$-Matrix Model}
\label{s-mehta-model}

Suppose $\tr S(A) = \tr [{c} A_1 A_2 + {g \over 4} (A_1^4 + A_2^4)
]$. The matrix integrals converge and the cyclic coupling tensors
are $S^{1111}=S^{2222}={g \over 4}$ and $S^{12}= S^{21} = {c \over
2}$. The LE for each $I$ are
    \beq
    c G_{2I} + g G_{111I} = \d^{I_1 1 I_2}_I G_{I_1} G_{I_2} &{\rm and}&
    c G_{1I} + g G_{222I} = \d^{I_1 2 I_2}_I G_{I_1} G_{I_2}.
    \eeq
Since the action is an $m=4^{\rm th}$ order polynomial, the LE do
not fix $G_i, G_{ij}$. They determine an infinite number of higher
rank correlations in terms of these, but also leave an infinite
number undetermined. For $I = \emptyset$ the two LE give $G_{111} =
-{c \over g} G_2$ and $G_{222} = -{c \over g} G_1$. The other
rank-$3$ correlations $G_{112}, G_{122} $ are left undetermined. For
$I=i_1$, the LE determine $4$ of $6$ correlations leaving $G_{1122}$
and $G_{1212}$ undetermined:
    \beq
    G_{1111} = G_{2222} = \ov{g} (1-c G_{12}),~~ G_{1112} = -{c \over
    g} G_{22},~~ G_{1222} = -{c \over g} G_{11}.
    \eeq
For $I=i_1 i_2$, the LE are
    \beq
    c G_{2 i_1 i_2} + g G_{111 i_1 i_2} = \d^1_{i_2} G_{i_1} +
        \d^1_{i_1} G_{i_2} &{\rm and}&
    c G_{1 i_1 i_2} + gG_{222 i_1 i_2} = \d^2_{i_2} G_{i_1} +
        \d^2_{i_1} G_{i_2}.
    \eeq
They determine $6$ of the $8$ rank-$5$ correlations in terms of
lower rank ones
    \beq
    G_{11111} = \ov{g}(2G_1 - c G_{112}),
    &G_{11112} = \ov{g}(G_2 - c G_{122}),&
    G_{11122} = {c^2 \over g^2} G_1, \cr
    G_{22222} = \ov{g}(2G_2 - c G_{122})
    &G_{12222} = \ov{g}(G_1 - c G_{112}),&
    G_{11222} = {c^2 \over g^2} G_2,
    \eeq
while leaving $G_{12121}$ and $G_{21212}$ undetermined. In this
manner, by choosing longer words $I$, we can fix an infinite number
of higher rank correlations in terms of lower rank ones, but at each
step a few correlations remain undetermined. The number of
undetermined correlators may be significantly reduced by the $A_1
\leftrightarrow A_2$ symmetry of $S(A)$ which implies $G_I = G_J$ if
$I$ can be obtained from $J$ by $1 \leftrightarrow 2$ and a cyclic
permutation. Notice that this is also a symmetry of the integration
measure. The same applies to the change of variables $A_1 \mapsto
-A_1, A_2 \mapsto -A_2$.

\subsubsection{Chern-Simons Model}
\label{s-cs-model}

The LE of the CS model $\tr S(A) = {2i\k \over 3} \eps^{ijk}\tr A_i
A_j A_k$ are
    \beq
    2 i \k \eps^{ijk} G_{Ijk} = \d^{I_1 i I_2}_I G_{I_1} G_{I_2}.
    \eeq
They leave rank-$1$ correlations $G_i$ undetermined $(m=3)$. For
$|I|=0$ and arbitrary $i$, the LE are $\eps^{ijk} G_{jk} =0$ which
do not give any constraints not already implied by cyclic symmetry
of $G_{jk}$. Thus $G_{12}, G_{13}, G_{23}, G_{11}, G_{22}, G_{33}$
are all left undetermined. For $|I|=1$ with arbitrary $I = i_1$ and
$i$, the LE are $2i \k \eps^{jki} G_{jk i_1} = \d^i_{i_1}$. From $9$
possible (complex) equations we get only $1$ independent condition
after accounting for cyclicity and hermiticity: the imaginary part
of
    \beq
    G_{123} - G_{132} = \ov{2 i \k}.
    \eeq
This allows us to fix only one parameter in the $c(3,\La = 3)=11$
dimensional space of $3^{\rm rd}$ rank cyclic hermitian tensors (see
appendix \ref{a-cyclic-tensors}). For $I=i_1 i_2$ and $i$ arbitrary,
the LE are
    \beq
    2 i \k \eps^{ijk} G_{i_1 i_2 jk} = \d^i_{i_2} G_{i_1} + \d^i_{i_1}
        G_{i_2}.
    \eeq
Of the $27$ possible equations, there are actually only $9$
independent ones that do not follow from cyclicity\footnote{The fact
that many of the loop equations are not independent of each other
indicates there are vector fields $v^I_i$ for which both sides of
\ref{e-loop-eqns-for-general-vector-field} vanish identically.}.
Three `homogeneous' ones $G_{1212} = G_{1122}$, $G_{1313} =
G_{1133}$, $G_{2323} = G_{2233}$ and six `inhomogeneous' ones
    \beq
    2i \k (G_{1123} - G_{1213}) = G_1, && 2i \k (G_{1213} -
        G_{1132}) = G_1 \cr
    2i \k (G_{1223} - G_{1232}) = G_2, && 2i \k (G_{1232} -
        G_{1322}) = G_2 \cr
    2i \k (G_{1323} - G_{1332}) = G_3, && 2i \k (G_{1233} -
        G_{1323}) = G_3.
    \eeq
Nevertheless, these conditions are not enough to fix the
$c(4,\La=3)=24$ independent cyclic and hermitian $4^{\rm th}$-rank
tensors (see appendix \ref{a-cyclic-tensors}). This underdetermined
nature of the LE persists for correlations of higher rank. Notice
also that by $A_1 \to A_2 \to A_3 \to A_1$ symmetry of the action
and measure, we have $G_1 = G_2 = G_3$ etc, but this is not a
consequence of the LE and still leaves the common value of these
undetermined.

\subsection{Additional equations for gluon correlations}
\label{s-additional-eqns}

Are there more equations satisfied by $G_I$ that will lessen the
underdeterminacy of the LE? In going from finite-$N$ SDE
(\ref{e-finite-N-SD-eq}) to large-$N$ LE (\ref{e-loop-eqns}), we
overlooked the possibility that both LHS and RHS of
(\ref{e-loop-eqns-for-general-vector-field}) may vanish for some
$v$. In other words, $A_i \to A_i + v^I_i A_I$ may leave the
(factorized expectation value of) action and measure simultaneously
invariant at leading order as $N \to \infty$. For such $v^I_i$ the
${\cal O}(N^0)$ terms in (\ref{e-finite-N-SD-eq}) identically vanish
and the ${\cal O}(1/N^2)$ terms constitute the leading large-$N$
SDE. Denote
    \beq
    \bra \Phi_I \ket = G_I + {G_I^{(2)} \over N^2} + {G_I^{(4)} \over N^4}
        + \ldots; ~~~~
    \bra \Phi_{I_1} \Phi_{I_2} \ket = G_{I_1} G_{I_2} + {G_{I_1;
        I_2}^{(2)} \over N^2} + {G_{I_1; I_2}^{(4)} \over N^4} +
        \ldots
    \eeq
Then the ${\cal O}(1/N^2)$ terms in (\ref{e-finite-N-SD-eq}) become
    \beq
    v^I_i S^{J_1 i J_2} G^{(2)}_{J_1 I J_2} = v^I_i \d_I^{I_1 i I_2} G^{(2)}_{I_1;
    I_2}  + v^I_i \sum_{p=1}^n \d_{K_p}^{L_p i M_p} G_{L_p I M_p}
    ~~~ \forall~~ v, ~K_p {\rm ~~and~~} n=1,2,\ldots
    \label{e-1-ov-Nsq-SDE}
    \eeq
Unfortunately, (\ref{e-1-ov-Nsq-SDE}) involve not just the $G_I$ but
also $1/N^2$ corrections to single and double-trace correlations.
Thus, an attempt to ameliorate the underdetermined nature of the LE
seems to open a new can of worms. However, in keeping with the
spirit of the large-$N$ limit as an approximation where we retain
only the leading large-$N$ contribution to all quantities, it seems
reasonable to ignore the $G^{(2)}_{\cdots}$ terms and consider
    \beq
    \sum_{p=1}^n v^I_i \d_{K_p}^{L_p i M_p} G_{L_p I M_p} =0
        \Leftrightarrow
    \sum_{p=1}^n v^I_i L_I^i G_{K_p} = 0  \Leftrightarrow
    \sum_{p=1}^n L_v G_{K_p} = 0
    \eeq
At first, these equations seem universal, they do not involve the
coupling tensors $S^I$ at all! However, for generic $v$, these are
$1/N^2$ contributions to the SDE and should be ignored in the
large-$N$ limit. But if $v^I_i$ are such that both RHS and LHS of
(\ref{e-loop-eqns-for-general-vector-field}) vanish identically,
then these become the leading large-$N$ SDE. Thus, these equations
are {\em not} universal, since they must be enforced only for those
$v^I_i$ for which the leading change in action and measure vanish
identically. To summarize, the additional equations are
    \beq
    \sum_{p=1}^n L_v G_{K_p} =0 && \forall
    ~~ K_1,\cdots,K_n {\rm ~~and~~} n=1,2,3 \ldots \cr && {\rm ~~and~~ all~~}
        v^I_i {\rm ~~such~that~~}
    v^I_i S^{J_1 i J_2} G_{J_1 I J_2} = v^I_i \d_I^{I_1 i I_2} G_{I_1} G_{I_2}
    =0.
    \label{e-aditional-eq}
    \eeq
Are there any such additional equations? This is related to whether
there are any transformations that leave both action and measure
invariant at leading order as $N \to \infty$. We exhibited several
such discrete transformations in sections
\ref{s-non-uniq-of-loop-eq-solns}, \ref{s-mehta-model} and
\ref{s-cs-model}. BRST transformations of gauge fixed Yang-Mills
theory are also of this sort and lead to Ward or Slavnov-Taylor
identities. Are the LE (\ref{e-loop-eqns}) consistent with the
additional equations (\ref{e-aditional-eq})? This would vindicate
our throwing away the subleading $G^{(2)}_{\cdots}$ terms in
(\ref{e-1-ov-Nsq-SDE}). If so, do the LE (\ref{e-loop-eqns})
together with (\ref{e-aditional-eq}) determine the $G_I$, or do we
need yet more conditions? We postpone investigation of these very
interesting issues and focus on the LE in the rest of this paper.

\subsection{Loop equation in terms of left annihilation and concatenation}
\label{s-loop-eqns-left-ann-conc}

Define the generating series of gluon correlations by the formal sum
$G(\xi) = G_I \xi^I$. Here, $\xi^i, 1 \leq i \leq \La$ are
non-commuting variables that can be thought of as sources, and
$\xi^{i_1 \cdots i_n} = \xi^{i_1} \cdots \xi^{i_n}$. If they did
commute, the generating series would only contain information about
the symmetric correlations. But since $G_{i_1 \cdots i_n}$ are not
symmetric in general (only cyclically symmetric), there is no
relation between $\xi^i \xi^j$ and $\xi^j \xi^i$. Define the
concatenation product $conc$ by
    \beq
    \xi^I \xi^J = \xi^{IJ} {\rm ~~or~~} F(\xi) G(\xi) = F_I G_J \xi^{IJ}
        ~~\Rightarrow~~   (FG)_K = \d_K^{IJ} F_I G_J.
    \label{e-concat-prod}
    \eeq
For example\footnote{Note that concatenation of cyclically symmetric
tensors is not cyclically symmetric in general.},
    \beq
    (FG)_0 = F_0 G_0; ~~~ (FG)_i = F_i G_0 + F_0 G_i; ~~~ (FG)_{ij} = F_0
    G_{ij} + F_i G_j + F_{ij} G_0; {\rm ~~~ etc.}
    \eeq
In terms of $conc$, the RHS of (\ref{e-loop-eqns}) becomes
$\delta^{I_1 i I_2}_I G_{I_1} G_{I_2} = [G(\xi) \xi_i G(\xi)]_I$.
Also define left annihilation\footnote{Left annihilation does not
preserve cyclic symmetry of tensors in general.}
    \beq
    D_j \xi^{i_1 \cdots i_n} = \d^{i_1}_j \xi^{i_2 \cdots i_n}.
    \label{e-left-annihilation}
    \eeq
$D_j$ eliminates the left most source if $i_1 = j$ and returns zero
otherwise. In terms of coefficients,
    \beq
    [D_j G]_I = G_{jI}, ~~~
    [D_{j_n} \cdots D_{j_1} G]_I  = G_{j_1 \cdots j_n I},
    \eeq
so that $G_{JI} = [D_{\bar J} G]_I$. The LE (\ref{e-loop-eqns}), one
for each $i$, can be written as
 \beq
    \sum_{n \geq 0} (n+1) S^{j_1 \cdots j_n i} D_{j_n} \cdots D_{j_1} G(\xi) =  G(\xi)
        \xi^i G(\xi)  {\rm ~~~ or ~~~} {\cal S}^i G(\xi ) = G(\xi)
        \xi^i G(\xi).
    \label{e-loop-eqns-in-terms-of-left-ann}
 \eeq
We used cyclicity of $S^I,G_I$ in deriving this. Thus, the LE
involve left annihilation and $conc$ product. The LHS of
(\ref{e-loop-eqns-in-terms-of-left-ann}) defines the action
dependent operator
    \beq
    {\cal S}^i = \sum_{n \geq 0} (n+1) S^{j_1 \cdots j_n i} D_{j_n} \cdots
    D_{j_1}.
    \eeq
At first glance, the LE (\ref{e-loop-eqns-in-terms-of-left-ann})
look like quadratically non-linear PDEs whose order is one less than
that of the action polynomial. However, concatenation in the
universal term on the RHS is non-commutative since sources $\xi^i$
do not commute. Further, left annihilation does not satisfy the
Leibnitz rule with respect to concatenation, i.e. $D_j$ are {\em
not} derivations of $conc$. This `mismatch-match' between product
and annihilation make the LE difficult to solve. It turns out there
is another natural product between gluon correlation tensors, the
shuffle product, with respect to which left annihilation satisfies
the Leibnitz rule. We try to exploit the interplay between $conc$,
shuffle and their derivations to find an approximation method to
solve the LE.

\subsection{Shuffle multiplication from products of
Wilson loop expectation values} \label{s-sh-prod-from-wilson-loops}

Here we obtain the shuffle product of gluon correlations induced by
expectation values of products of Wilson loops. The expectation
value of the Wilson loop $F(\gamma)$ is a complex-valued
gauge-invariant function on the space of loops $\gamma : S^1 \to M$,
where $M$ is space-time. If $A_\nu(x)$ denotes the components of a
gauge field $1$-form valued in the Lie algebra of hermitian
matrices, we define the path ordered exponent
    \beq
    F(\gamma) = \ov{N} \tr {\cal P} \exp\bigg[i \int_0^1 A_\nu(x) {d x^\nu \over
    ds} ds \bigg].
    \eeq
Parameterized loops on $M$ are denoted $x^\nu(s)$. Wilson loops are
typical functions on loop-space and their expectation values can be
expanded in iterated integrals of gluon correlations
    \beq
    \bra F(\g) \ket &=& \sum_{m=0}^\infty i^m \int_{0 \leq s_1 \leq \cdots \leq s_m \leq 1}
    \bra \ov{N} \tr A_{\nu_1}(x(s_1)) \cdots A_{\nu_m}(x(s_m)) \ket {d x^{\nu_1} \over
        ds_1} \cdots {d x^{\nu_m} \over ds_m} ds_1 \cdots ds_m \cr
    &=& \sum_{m=0}^\infty i^m \int_{0 \leq s_1 \leq \cdots \leq s_m \leq
    1} F_{\nu_1 \cdots \nu_m}(x(s_1), \cdots, x(s_m)) {d x^{\nu_1} \over
        ds_1} \cdots {d x^{\nu_m} \over ds_m} ds_1 \cdots ds_m
    \eeq
where the gluon correlation tensors associated to $F(\g)$ are
    \beq
    F_{\nu_1 \cdots \nu_m}(x(s_1), \cdots, x(s_m))
        = \bra \ov{N} \tr A_{\nu_1}(x(s_1)) \cdots A_{\nu_m}(x(s_m))
        \ket.
    \eeq
The point-wise commutative product of functions on loop-space is
defined as $(F G)(\gamma) = F(\gamma) G(\gamma)$. Taking
expectation-values and working in the large-$N$ limit, where
correlations factorize, we get
    \beq
    \bra (FG)(\g) \ket = \bra F(\gamma) G(\gamma) \ket
    = \bra F(\gamma) \ket \bra G(\gamma) \ket + {\cal O}(\ov{N^2}).
    \eeq
We may expand the LHS in correlation functions associated to the
Wilson loop $(FG)(\gamma)$. We call these $(F \circ G)_{\rho_1
\cdots \rho_p}(x(u_1) \cdots x(u_p))$. They are defined as
    \beq
    \bra (FG)(\g) \ket = \sum_{p=0}^\infty i^p \int_{0 \leq u_1 \leq \cdots \leq u_p \leq 1}
        (F \circ G)_{\rho_1 \cdots \rho_p}(x(u_1) \cdots x(u_p)) {d x^{\rho_1} \over
        du_1} \cdots {d x^{\rho_p} \over du_p} du_1 \cdots du_p.
    \eeq
Meanwhile, the expansion of the RHS reads
    \beq
    \bra F(\gamma) \ket \bra G(\gamma) \ket &=& \sum_{m,n=0}^\infty
    i^{m+n}
    \int_{\stackrel{0\leq s_1 \leq \cdots s_m \leq 1}{0\leq t_1 \leq \cdots t_n \leq 1}}
        F_{\nu_1 \cdots \nu_m}(x(s_1), \cdots, x(s_m))
        G_{\mu_1 \cdots \mu_n}(x(t_1), \cdots, x(t_n)) \cr
    && \times {d x^{\nu_1} \over ds_1} \cdots {d x^{\nu_m} \over ds_m}
        {d x^{\mu_1} \over dt_1} \cdots {d x^{\mu_n} \over dt_n}
        ds_1 \cdots ds_m dt_1 \cdots dt_n.
    \eeq
To make this look like the expansion of the LHS, we collect terms
with a common sum $n+m = p$ and then sum from $p = 0$ to $\infty$.
Moreover, we must relabel the $\nu$'s and $\mu$'s  as $\rho$'s and
the $s$'s and $t$'s as $u$'s. We must allow every possible
relabeling that preserves the order among the $s$'s and $t$'s. When
this is done, we read off the relation between the gluon
correlations associated to the Wilson loop $(FG)(\g)$ and those
associated to $F(\g)$ and $G(\g)$
    \beq
    (F \circ G)_{\rho_1 \cdots \rho_p}(x(u_1) \cdots x(u_p)) &=&
    \sum_{m+n=p} \sum_{\stackrel{\s {\rm ~an~} (m,n) }{{\rm ~shuffle}}}
    F_{\rho_{\s^{-1}(1)} \cdots \rho_{\s^{-1}(m)}}(x(u_{\s^{-1}(1)}), \cdots, x(u_{\s^{-1}(m)}))
    \cr && \times
    G_{\rho_{\s^{-1}(m+1)} \cdots \rho_{\s^{-1}(m+n)}}(x(u_{\s^{-1}(m+1)}),
        \cdots, x(u_{\s^{-1}(m+n)})).
    \eeq
An $(m,n)$ shuffle is a permutation of $m+n$ letters $(1,2, \cdots
,m+n)$ such that
    \beq
    \sigma^{-1}(1) < \cdots < \sigma^{-1}(m) {\rm ~and~}
    \sigma^{-1}(m+1) < \cdots < \sigma^{-1}(m+n).
    \eeq
For brevity, we combine the Lorentz $\mu$ and space-time $x^\mu$
indices into a single index $i$, then
    \beq
    (F \circ G)_{i_1 \cdots i_p} = \sum_{m+n =p}~ \sum_{\s {\rm ~an~} (m,n) {\rm ~shuffle}}
    F_{i_{\s^{-1}(1)}, \cdots i_{\s^{-1}(m)}} G_{i_{\s^{-1}(m+1)}, \cdots
    i_{\s^{-1}(m+n)}}.
    \eeq
The RHS is called the shuffle product ($sh$). It is commutative. A
compact notation for $sh$ is
    \beq
    (F \circ G)_I = \sum_{I = J \sqcup K} F_J G_K.
    \label{e-sh-prod-of-gluon-corrlns}
    \eeq
The condition $I= J \sqcup K$ means that $J$ and $K$ are
complementary order-preserving sub-words of $I$. The operation $J
\sqcup K$ is a riffle-shuffle of two card packs $J$ and $K$. Some
examples are
    \beq
    [F \circ G]_i &=& F_i G_0 + F_0 G_i; ~~~~~~~
    [F \circ G]_{i j} = F_{i j} G_0 + F_i G_j
        + F_j G_i + F_0 G_{ij}; \cr
    [F \circ G]_{ijk} &=& F_{ijk} G_0 + F_{ij} G_k + F_{ik} G_j + F_{jk}
        G_i \cr && + F_i G_{jk} + F_j G_{ik}
        + F_k G_{ij} + F_0 G_{ijk}; \cr
    [F \circ G]_{ijkl} &=& F_{ijkl} G_0 + F_{ijk} G_l + F_{ijl} G_k
        + F_{ikl} G_j + F_{jkl} G_i \cr && + F_{ij} G_{kl} + F_{ik} G_{jl} +
        F_{il} G_{jk} + F_{jk} G_{il} + F_{jl} G_{ik} + F_{kl} G_{ij} \cr && +
        F_i G_{jkl} + F_j G_{ikl} + F_{k} G_{ijl} + F_l G_{ijk} + F_0
        G_{ijkl}.
    \eeq
We notice two properties of $sh$. If $F_I$ and $G_J$ are cyclically
symmetric for all $I$ and $J$, then so is $(F \circ G)_K$ for all
$K$. To see why this is true in general, observe that $(F \circ
G)_K$ is the expectation value of the trace of a product of gluon
fields, and the trace makes it cyclically symmetric. Thus $sh$
preserves cyclicity of tensors. Moreover, we notice that if $F_I$
and $G_J$ satisfy the hermiticity properties $F_I^* = F_{\bar I},
G_J^* = G_{\bar J}$ for all $I,J$, then so does their shuffle
product
    \beq
    (F \circ G)_I^* = (F \circ G)_{\bar I} ~~~ ~ \forall~~ I.
    \eeq
This is a reflection of the relations\footnote{$\bar \g$ is the loop
$\g$ with opposite orientation.} $F(\g)^* = F(\bar \g)$ and
$(FG)^*(\g) = F^*(\g) G^*(\g) = (FG)(\bar \g)$ when the path-ordered
exponential is expanded out in iterated integrals.

The shuffle product allows us to reduce manipulations in the
commutative algebra of functions on the infinite dimensional space
$Loop(M)$ to operations on tensors on the finite dimensional space
$M$. More precisely, start with a manifold $M$, and denote the space
of $1$-forms on $M$ by $\La^1(M)$. Then consider the tensor algebra
$\cal T$ on $\La^1(M)$. The shuffle algebra is
    \beq
    Sh(M) = {\cal T}(\La^1(M)).
    \eeq
The shuffle algebra is a replacement for the algebra of functions on
$Loop(M)$. Let $\xi^{i_1}, \xi^{i_2}, \ldots$ be a basis for
$\La^1(M)$ (think of these as $dx^{i_1}, \ldots$), then an element
of the shuffle algebra is
    \beq
    G = \sum_n G_{i_1 \cdots i_n} \xi^{i_1} \otimes \cdots \otimes
           \xi^{i_n} \equiv \sum_n G_{i_1 \cdots i_n} \xi^{i_1 \cdots
           i_n},
    \eeq
and is to be regarded as a function on $Loop(M)$. A specific
collection of gluon correlations $\{ G_{i_1 \cdots i_n}
\}_{n=0}^\infty$ can encode the information contained in the
expectation value of a specific function $G(\g)$ on
$Loop(M)$\footnote{The map is not 1-1 since gluon correlations are
not gauge invariant in general, unlike Wilson loops. A way to deal
with this is to introduce ghosts. When LE are formulated in terms of
Wilson loops, gauge fixing and ghost contributions cancel out
\cite{makeenko-book}. But this is not the case if we work with
correlation tensors. Extension of this formalism to include ghosts
in matrix models will be treated in
\cite{gsk-ghost-shuffle-concat}.}. The shuffle product of basis
elements is
    \beq
    \xi^{i} \circ \xi^{j} = \xi^{ij} + \xi^{ji}; &&
    \xi^{ij} \circ \xi^{k} = \xi^{ijk} + \xi^{ikj}+ \xi^{kij}
    \eeq
and in general
    \beq
    \xi^{i_1 \cdots i_p} \circ \xi^{i_{p+1} \cdots i_{p+q}}
        = \sum_{\sigma {\rm ~a~} (p,q) \rm{~shuffle}}
        \xi^{i_{\sigma(1)} \cdots i_{\sigma(p+q)}}  {\rm ~~~~~or~~~}
    \xi^J \circ \xi^K = \d_I^{J \sqcup K} \xi^I.
    \eeq
To summarize, we have shown that the commutative point-wise product
of Wilson loops induces the commutative, cyclicity and hermiticity
preserving shuffle product of gluon correlations\footnote{This
construction generalizes to differential forms on $Loop(M)$, but we
do not use it in this paper.}.

\subsection{Derivations of shuffle and concatenation products}
\label{s-derivations-of-sh-conc}

Concatenation and shuffle combine to define a pair of dual
bialgebras on the vector space ${\rm span}(\xi^I)$ (see appendices
\ref{a-conc-sh-and-coprods} and \ref{a-bialgebra-structure}).
Derivations of concatenation and shuffle play a central role in this
paper. Recall that the LE (\ref{e-loop-eqns-in-terms-of-left-ann})
involved left annihilation $D_i$ defined in
(\ref{e-left-annihilation}). We show that $D_i$ is a derivation of
$sh$ i.e. it satisfies the Leibnitz rule
 \beq
    D_i (F \circ G) = (D_i F) \circ G + F
    \circ (D_i G).
 \eeq
The proof is by explicit calculation $[D_i (F \circ G)]_I = [F \circ
G]_{iI} = \sum_{I_1 \sqcup I_2 = iI} F_{I_1} G_{I_2}$. Now either $i
\in I_1$ or $i \in I_2$, so
 \beq
    [D_i (F \circ G)]_I &=& \sum_{I_1 \sqcup I_2 = I} F_{i I_1}
        G_{I_2}+ \sum_{I_1 \sqcup I_2 = I} F_{I_1}
        G_{i I_2}
    = \sum_{I_1 \sqcup I_2 = I} [D_i F]_{I_1}
        G_{I_2} + \sum_{I_1 \sqcup I_2 = I} F_{I_1} [D_i
        G]_{I_2} \cr
    &=& [(D_i F) \circ G]_I + [F \circ (D_i G)]_I.
 \eeq
Full annihilation\footnote{{\em Not} the cyclic gradient. The cyclic
gradient is $\delta_i \xi^I = \delta^I_{I_1 i I_2} \xi^{I_2 I_1}$
and is {\em not} a derivation of concatenation.} $\bfD_j$ is a
democratic version of left annihilation. It is defined as
 \beq
    \bfD_j \xi^I = \delta^I_{I_1 j I_2} \xi^{I_1 I_2} {\rm ~~~~ and ~~~}
    [\bfD_j F]_I = \delta_I^{I_1 I_2} F_{I_1 j I_2}.
    \label{e-full-annihilation}
 \eeq
$\bfD_j$ does not preserve cyclic symmetry of tensors. However,
$\bfD_j$ is a derivation of $conc$,
 \beq
    \bfD_j (F G) = (\bfD_j F) G + F (\bfD_j G).
 \eeq
To see this, begin with the LHS $[\bfD_j(F G)]_I = \delta_I^{I_1
I_2} (F G)_{I_1 j I_2}$,
 \beq
    [\bfD_j(F G)]_I
        = \delta_I^{I_1 I_2} \delta_{I_1 j I_2}^{K_1 K_2}
            F_{K_1} G_{K_2}
        = \delta_I^{L_1 L_2 L_3} F_{L_1 j L_2} G_{L_3} +
            \delta_I^{L_1 L_2 L_3} F_{L_1} G_{L_2 j L_3}.
    \eeq
On the other hand,
    \beq
    [(\bfD_j F) G]_I = \d_I^{I_1 I_2} (\bfD_j F)_{I_1} G_{I_2}
        = \d_{I}^{I_1 I_2} \d_{I_1}^{J_1 J_2} F_{J_1 j J_2} G_{I_2}
        = \d_I^{L_1 L_2 L_3} F_{L_1 j L_2} G_{L_3}.
    \eeq
Thus
    \beq
    [(\bfD_j F) G]_I + [F (\bfD_j G)]_I =
        \delta_I^{L_1 L_2 L_3} F_{L_1 j L_2} G_{L_3} +
            \delta_I^{L_1 L_2 L_3} F_{L_1} G_{L_2 j L_3} =
            [\bfD_j(F  G)]_I.
    \eeq
The commutator of derivations is a derivation irrespective of
whether the product is commutative or not. This is analogous to the
Lie bracket of vector fields being a vector field on a manifold. For
example, merely using the fact that $D_i$ is a derivation of $sh =
\circ$, it is easy to show that
    \beq
        [D_i, D_j] (F \circ G) = ([D_i,D_j]F) \circ G + F \circ ([D_i, D_j]G).
    \eeq
It follows that iterated commutators of derivations (e.g.
$[D_i,[D_j,D_k]]$) are also derivations. On the other hand, products
of left annihilation operators are not derivations of the shuffle
algebra. For e.g. $D_i D_j = D_{ij}$ is not a derivation of $sh$.
This is analogous to the product of vector fields not being a vector
field. Furthermore, left annihilation operators with a single index
$D_i$ do not form a Lie algebra by themselves. The commutator $[D_i,
D_j] = D_{ij} - D_{ji}$ is not a linear combination of $D_k$'s.
However, by construction, the vector space spanned by the set of all
iterated commutators of left annihilation operators $D_i, [D_i,D_j],
[D_i,[D_j,D_k]], \cdots$ forms a Lie algebra, the Lie algebra of
derivations of the shuffle product. This is the free Lie algebra. It
is analogous to the Lie algebra of left invariant vector fields on a
Lie group. Here, the role of the Lie group is played by the free
group on $\La$ generators.

\subsection{Derivation property of terms in
Yang-Mills action} \label{s-derivation-property}

The action-dependent linear term ${\cal S}^i G(\xi)$ in the LE
(\ref{e-loop-eqns-in-terms-of-left-ann}) is a sum of products of
left annihilation operators ${\cal S}^i = \sum_{n \geq 0} (n+1)
S^{j_1 \cdots j_n i} D_{j_n} \cdots D_{j_1}$. Suppose coupling
tensors $S^I$ are such that ${\cal S}^i$ is a linear sum of iterated
commutators of left annihilation operators,
    \beq
    {\cal S}^i = C^{ij} D_j + C^{ijk} [D_j,D_k] + C^{ijkl}
    [[D_j,D_k],D_l] + \cdots
    \label{e-lin-comb-iterated-commutators}
    \eeq
Then ${\cal S}^i$ is a derivation of shuffle. Of what practical use
is this property? The LE (\ref{e-loop-eqns-in-terms-of-left-ann})
are quadratically non-linear in $conc$, but involve left
annihilation, which is a derivation of $sh$. In section
\ref{s-approx-meth-multi-mat} we introduce an approximation scheme
where $conc$ is expanded around $sh$. The main simplification for
matrix models having the derivation property is that their LE can be
turned into (an infinite system of) {\em linear} PDEs at $0^{\rm
th}$ order in this approximation. This is not the case for matrix
models without the derivation property.

Among $1$-matrix models, the only one with this property is the
Gaussian $\tr S(A) = \ov{2\a} \tr A^2$ for which ${\cal S} = \ov{\a}
D$. For $\La =1$, there is only one left annihilation operator, and
all its iterated commutators vanish. Multi-matrix models provide
non-trivial examples. It is remarkable that the gluonic terms in the
Yang-Mills action (\ref{e-gauge-fixed-ym-action}) quadratic in
momentum, linear in momentum and independent of momentum each
separately has this derivation property\footnote{See
\cite{gsk-ghost-shuffle-concat} for the corresponding property after
inclusion of ghosts.}. These terms can be written as $\tr C^{ij} A_i
A_j, \tr C^{ijk} A_i [A_j, A_k]$ and $\tr [A_i,A_j][A_k,A_l] g^{ik}
g^{jl}$ for appropriate tensors $C^{ij}, C^{ijk}, g^{ij}$. Moreover,
the zero momentum limits of the Gaussian, Chern-Simons and
Yang-Mills matrix field theories all have this derivation property.
They correspond to the simplest non-vanishing choices for the
tensors $C^{ij}, C^{ijk}, C^{ijkl}$ in
(\ref{e-lin-comb-iterated-commutators}). In fact, this property also
extends to the corresponding matrix field theories but we do not
address that here.

{\flushleft \bf Gaussian:} The Gaussian multi-matrix model $\tr S(A)
= \half \tr C^{ij} A_i A_j$ has real-symmetric covariance
$C^{ij}=C^{ji}$. $S^{ij} = \half C^{ij}$ is cyclically symmetric and
also satisfies $(S^{ij})^* = S^{ji}$ so that all correlations
satisfy $G_I^* = G_{\bar I}$. We get ${\cal S}^i = 2 S^{ij}D_j =
C^{ij} D_j$, which is a linear combination of left annihilation
operators and therefore a derivation of $sh$. The LE are
    \beq
    C^{ij} D_j G(\xi) = G(\xi) \xi^i G(\xi).
    \eeq
{\flushleft \bf Chern-Simons:} For at least three matrices $(\La
\geq 3)$, the CS type of matrix model has action ${2 i \k \over 3}
\tr C^{ijk} A_i [A_j, A_k]$ where $C^{ijk}$ is any tensor which is
anti-symmetric under interchange of any pair of indices. The part of
$C^{ijk}$ that is symmetric under interchange of a pair of indices
does not contribute on account of antisymmetry of the commutator.
The action can also be written as $\tr S(A) = {2 i \k \over 3} \tr
\tl C^{ijk} A_i A_j A_k$ where $\tl C^{ijk} = C^{ijk} - C^{ikj}$.
The particular case of zero momentum 3d CS gauge theory results from
the choice $\La =3$, $\tl C^{ijk} = \eps^{ijk}$ (the Levi-Civita
symbol), and integer-valued coupling constant $4 \pi \k$. More
importantly, terms in the Yang-Mills action
(\ref{e-gauge-fixed-ym-action}) linear in momentum are of this form.
Irrespective of its field theoretic origin, $S^{ijk} = (2 i \k/3)
\tl C^{ijk}$ is cyclically symmetric since $\tl C^{kij} = (-1)^2 \tl
C^{ijk}$. Moreover, $(S^{ijk})^* = S^{kji}$ so that $G_I^* = G_{\bar
I}$. Now ${\cal S}^i$ is a linear combination of commutators of left
annihilation operators:
    \beq
    {\cal S}^i = 2 i \k \tl C^{ijk} D_k D_j = i \k \{\tl C^{ijk}
    D_k D_j - \tl C^{ikj} D_k D_j \} = i \k \tl C^{ijk} [D_k, D_j]
    \eeq
and therefore a derivation of $sh$. The `Chern-Simons' loop
equations are
    \beq
        i \k \tl C^{ijk} [D_k, D_j] G(\xi) = G(\xi) \xi^i G(\xi).
    \eeq

{\flushleft \bf Yang-Mills:} For $\La \geq 2$, the zero momentum
limit of Yang-Mills theory has action ($\a = g^2$)
    \beq
    \tr S(A) = - \ov{4 \a} \tr [A_i,A_j][A_k,A_l] g^{ik} g^{jl},
    \eeq
where $g^{ij} = g^{ji}$ is the inverse metric, it is a real
symmetric matrix. The action is rewritten as
    \beq
    \tr S(A) = {-1 \over 2 \a} \tr (g^{ik} g^{jl} - g^{il} g^{jk}) A_{ijkl}
        = {-1 \over 4 \a} \tr[(2 g^{ik} g^{jl} - g^{il} g^{jk} - g^{ij} g^{kl}) A_{ijkl}]
    \eeq
so that $S^{ijkl} = -\ov{4 \a} (2 g^{ik} g^{jl} - g^{il} g^{jk} -
g^{ij} g^{kl})$ is cyclically symmetric. Moreover, $S^{ijkl} =
(S^{lkji})^* = S^{lkji}$ follows since $g^{ij}$ is real symmetric.
Then the differential operator ${\cal S}^i = (3+1) S^{ijkl} D_l D_k
D_j$
    \beq
    {\cal S}^i =  -\ov{\a} g^{ik} g^{jl}
        (D_l D_k D_j - D_k D_l D_j + D_l D_k D_j - D_l D_j D_k)
    = -\ov{\a} g^{ik} g^{jl} [D_j,[D_k,D_l]]
    \eeq
is a linear combination of iterated commutators of derivations and
hence a derivation of the shuffle product. The Yang-Mills LE are
thus
    \beq
        - \ov{\a} g^{ik} g^{jl} [D_j,[D_k,D_l]] G(\xi) = G(\xi) \xi^i
        G(\xi).
    \eeq
On the other hand, {\em most matrix models do not have this
derivation property}. For example, consider the popular
\cite{kazakov-marshakov} two matrix model $\tr S(A_1,A_2) =
\tr[A_1^4 + A_2^4 + 2 A_1 A_2]$. Here, ${\cal S}^1 = 2 D_2 + 4
D_1^3$ and ${\cal S}^2 = 2 D_1 + 4 D_2^3$ are not linear
combinations of iterated commutators of $D_i$ and do not define
derivations of the shuffle algebra.

\section{Approximation method for one-matrix models}
\label{s-approx-meth-one-mat}

The LE of a $1$-matrix model (\ref{e-1-mat-loop-eq}) with $m^{\rm
th}$ order polynomial action
    \beq
    \sum_{l=1}^m l~ S_l D^{l-1} G(\xi) = G(\xi) \xi G(\xi).
    \eeq
can be written in terms of left annihilation\footnote{The $1$-matrix
left annihilation operator, $D \xi^n = \xi^{n-1}$ is {\em not} the
same as the usual derivative of calculus.} $D$. Concatenation, which
appears on the RHS is the usual product of calculus. But $D$
satisfies the Leibnitz rule with respect to $sh$, not $conc$. So
this is not a differential equation. We develop approximation
methods to solve these LE either by expanding $conc$ around $sh$ or
by expanding $D$ around full annihilation $\bfD$
(\ref{e-full-annihilation}), which is a derivation of $conc$. Both
these turn the LE into linear ODEs at each order of the expansion.

\subsection{Shuffle, concatenation and their derivations}
\label{s-1-mat-sh-conc-derivations}

We give the $1$-matrix versions of $conc$, $sh$ and their
derivations by specialization from sections
\ref{s-loop-eqns-left-ann-conc} and
\ref{s-sh-prod-from-wilson-loops}. Then we define $q$-deformed
products and derivations that we use to solve the LE approximately.
Suppose $F(\xi) = \sum_{n \geq 0} F_n \xi^n$ etc. $Conc = *_1$ is
the usual product of calculus\footnote{We also denote $conc = *_1$
by juxtaposition.},
    \beq
    \xi^p *_1 \xi^q = \xi^{p+q}    ~~~{\rm or ~~~}
        (F *_1 G)_n = \sum_{r=0}^n F_r G_{n-r}
    \eeq
while shuffle $= *_0$ (previously denoted $\circ$) is,
    \beq
    \xi^p *_0 \xi^q = {p+q \choose p} \xi^{p+q}, {\rm ~~~ or ~~~}
     (F *_0 G)_n = \sum_{r=0}^n  {n \choose r} F_r G_{n-r}
    \eeq
For example $\xi *_0 \xi = 2 \xi^2$. Both are commutative. The
notation anticipates $*_q$ that interpolates between $sh$ ($q=0$)
and $conc$ ($q=1$). We also define $1$-matrix analogs of left and
full annihilation and name them in anticipation of $q$-annihilation
$D_q$. Left annihilation $D_0 \xi^n = \xi^{n-1}$ is the $1$-matrix
version of $D_i$ defined in (\ref{e-left-annihilation}). $D_0$ is a
derivation of shuffle
    \beq
    (D_0 F)_n = F_{n+1}, ~~~
    (D_0 (F *_0 G))_n = ((D_0 F) *_0 G)_n + (F *_0
        (D_0 G))_n.
    \eeq
Full annihilation $D_1 \xi^n = n \xi^{n-1}$ is the same as the usual
derivative of calculus. It is the $1$-matrix version of $\bfD_i$
defined in (\ref{e-full-annihilation}). $D_1$ is a derivation of
$conc$,
    \beq
    [D_1 F]_n = (n+1) F_{n+1}, ~~~~ (D_1 (F *_1 G))_n
    = ((D_1 F) *_1 G)_n + (F *_1 (D_1 G))_n.
    \eeq
This follows from the easily verified formula
    \beq
    (n+1) \sum_{r=0}^{n+1} F_r G_{n+1-r} = \sum_{r=0}^n (r+1)
    F_{r+1} G_{n-r} + \sum_{r=0}^n (n-r+1) F_r G_{n-r+1}.
    \eeq

\subsection{$q$-Deformed product}
\label{s-1-mat-q-deformed-prod}

The $q$-product interpolates between $conc$ $(q = 1)$ and $sh$ $(q =
0)$\footnote{The quantity $1-q$ often occurs in formulae, so we call
it $p = 1-q$.}
    \beq
    (F *_q G)_n = \sum_{r=0}^n  {n \choose r}_{1-q} F_r G_{n-r}.
    \label{e-def-q-prod-one-mat}
    \eeq
It is associative and commutative for $0 \leq q \leq 1$. The
$q$-binomial coefficients or Gauss binomials ${n \choose r}_q$ are
polynomials in $q$ with non-negative coefficients. They reduce to
unity for $q = 0$ and to the usual binomial coefficients when $q
=1$. To obtain their properties let $yx = q xy$. Then
    \beq
    (x + y)^n = \sum_{r=0}^n {n \choose r}_q x^{n-r} y^r.
    \label{e-def-gauss-binomials}
    \eeq
The first three Gauss binomials are
    \beq
    {n \choose 0}_q &=& 1, ~~~~~
    {n \choose 1}_q = 1 + q + q^2 + \cdots + q^{n-1}, \cr
    {n \choose 2}_q &=& \left\{
       \begin{array}{ll}
       (1+ q^2 + q^4 + \cdots + q^{n-2})(1 + q + q^2 + \cdots q^{n-2}), & \hbox{if $n$ is even;} \\
       (1+ q^2 + q^4 + \cdots + q^{n-3})(1+q + q^2 + \cdots + q^{n-1}) , & \hbox{if $n$ is odd.}
                          \end{array} \right.
    \label{e-eg-q-gauss-binomials}
    \eeq
The $q$-Pascal relation is got by multiplying $(x+y)^{n-1}$ by
$(x+y)$ either from the right or left:
    \beq
    {n \choose r}_q  &=&  q^r  {n -1 \choose r}_q + {n-1 \choose
        r-1}_q \cr
    {n \choose r}_q &=& {n-1 \choose r}_q + q^{n-r} {n-1 \choose
        r-1}_q.
    \eeq
Substituting the first in the second gives
    \beq
    {n \choose r}_q = {1-q^n \over 1-q^{n-r}} {n-1 \choose r}_q
    {\rm ~~~ for ~~~} 0 \leq r < n.
    \eeq
Iterating, we get
    \beq
    {n \choose r}_q = {(1-q^n) (1-q^{n-1}) \cdots (1-q^{n-r+1}) \over (1-q)
        (1-q^2) \cdots (1-q^r)}.
    \eeq
This can also be written as
    \beq
    {n \choose r}_q = {[n]_q! \over [r]_q! [n-r]_q!} {\rm ~~where~~}
    [n]_q! = [1]_q [2]_q \cdots [n]_q {\rm ~~and~~}
    [n]_q = {1-q^n \over 1-q}.
    \eeq
The symmetry ${n \choose r}_q = {n \choose n-r}_q$ is now manifest,
which guarantees commutativity of the $q$-product
(\ref{e-def-q-prod-one-mat}). Some examples of the $q$-product are
    \beq
    (F *_q G)_0 &=& F_0 G_0; ~~ (F *_q G)_1 = F_1 G_0 + F_0 G_1; ~~~~
    (F *_q G)_2 = F_0 G_2 + (1+p) F_1 G_1 + F_2 G_0; \cr
    (F *_q G)_3 &=& F_0 G_3 + (1+p+p^2) (F_1 G_2 + F_2 G_1) +
        F_3 G_0; \cr
    (F *_q G)_4 &=& F_0 G_4 + (1+p + p^2 + p^3) (F_1 G_3 + F_3 G_1) \cr &&
        + (1+p + 2 p^2 + p^3 + p^4) F_2 G_2 + F_4 G_0.
    \label{e-eg-q-shuffle-one-mat}
    \eeq
We expand the $q$-binomials around the ordinary binomial
coefficients ($q = 1$) in a Taylor series
    \beq
    {n \choose r}_q = {n \choose r}_{1} ~\bigg\{1 -
        {r (n-r) \over 2} p + {\cal O}(p^2)  \bigg\}.
    \eeq
Thus $*_q$ may be expanded around shuffle $*_0$
    \beq
    (F *_q G)_n &=& (F *_0 G)_n - {q \over 2} \sum_{r=0}^n {n \choose r}_1
        r F_r (n-r) G_{n-r} + \cdots \cr
    &=& (F *_0 G)_n - {q \over 2} \sum_{r=0}^n {n \choose r}_1 [\xi
        *_0 D_0 F(\xi)]_r [\xi *_0 D_0 G(\xi)]_{n-r} + \cdots   \cr
    (F *_q G)(\xi) &=& (F *_0 G)(\xi) - {q \over 2}~ \xi
    *_0 (D_0 F)(\xi) *_0 \xi *_0 (D_0 G)(\xi) + \cdots.
    \label{e-1-mat-q-series-for-q-prod}
    \eeq
Taking $q = 1$, and using commutativity of $*_0$, we get an
expansion for $conc$ in terms of $sh$
    \beq
    (F *_1 G)(\xi) = (F *_0 G)(\xi) - \half~ \xi *_0 \xi
    *_0  (D_0 F)(\xi) *_0 (D_0 G)(\xi) + \cdots.
    \eeq

\subsection{$q$-Deformed annihilation operator}
\label{s-1-mat-q-deformed-annihilation}

Recall from section \ref{s-1-mat-sh-conc-derivations} that left
annihilation $[D_0 F]_n = F_{n+1}$ and full annihilation $[D_1 F]_n
= F_{n+1}$. More generally, let
    \beq
    (D_q F)_n = [n+1]_q F_{n+1}
        = \bigg[ {q^{n+1} - 1 \over q -1} \bigg] F_{n+1}
        = \bigg[1 + q + q^2 + \cdots + q^n \bigg] F_{n+1}.
    \label{e-1-mat-q-annihilation}
    \eeq
$D_q$ reduces to left and full annihilation for $q=0$ and $q=1$.
However, $D_q$ is not a derivation of $*_q$ for $0<q<1$.
Fortunately, we don't seem to need that. More importantly, we expand
$D_q$ around $D_1$ in powers of $p = 1-q$. Denoting $conc$
reciprocal by usual division of calculus,
    \beq
    D_q F(\xi) = {F(q \xi) - F(\xi) \over (q-1)\xi}
        = \sum_{k=1}^\infty (-p \xi)^{k-1} \ov{k!} D_1^k F(\xi)
    \label{e-1-mat-q-annihilation-expanded-in-p}
    \eeq


\subsection{Gaussian one matrix model}
\label{s-1-mat-gaussian}

Now we apply this formalism to the simplest of matrix models, the
Gaussian  $1$-matrix model. We pick it as it is the only $1$-matrix
model with the derivation property. We show how expanding $conc$
around $sh$ and expanding $D_0$ around $D_1$, are used along with
the derivation property to turn the non-linear LE into linear ODEs
at each order in our approximation schemes. The resulting gluon
correlations are compared with the exact solution.

From (\ref{e-1-mat-loop-eq}), the LE for the Gaussian $1$ matrix
model with action $S = {1 \over 2 \a} \tr A^2$ are
    \beq
    D_0 Z(\xi)= \a \xi Z(\xi) *_1 \xi *_1 Z(\xi) {\rm ~~~or ~~~}
    G_{n+1} = \alpha \sum_{r+1+s=n, ~ r,s \geq 0} G_r G_s, ~~~~ n = 0,1,2,
    \cdots
    \label{e-gauss-1-mat-loop-eq}
    \eeq
with the boundary condition $G_0 = 1$. When the product is not
specified, it is taken to be the concatenation product $*_1$. In
this section, we call the generating function of moments $Z(\xi) =
\sum_n G_n \xi^n$. This is because we will expand $Z(\xi)$ in powers
of $q$, and the coefficients $Z_k(\xi)$ are not to be confused with
the moments $G_n$, which are coefficients in an expansion in powers
of $\xi$. Of course, $q$ is a bookkeeping device which is eventually
set to $1$.

\subsubsection{Exact solution}
\label{s-1-mat-gauss-exact-sol}

The loop equation for the Gaussian (\ref{e-gauss-1-mat-loop-eq}) may
be solved since it is a quadratic equation
    \beq
    {Z(\xi) - Z(0) \over \xi} = \alpha~ Z^2(\xi) \xi
    ~~\Rightarrow~~ \alpha \xi^2 Z^2 - Z +1 = 0.
    \eeq
The solution is
    \beq
    Z(\xi)  ~=~ {1 - \sqrt{1- 4 \alpha \xi^2} \over 2 \alpha \xi^2}
    ~~=~~ \sum \Gamma_{2n} \xi^{2n}
    \eeq
where $\Gamma_n$ are the moments. Define Catalan numbers $C_n$ by
    \beq
    \sum_{n=0}^\infty C_n x^n = {1 - \sqrt{1-4x} \over 2x}
    {\rm ~~with~~}
    C_n = {(2n)! \over n! (n+1)!} \sim {4^n \over \sqrt{n^3 \pi}} {\rm
    ~as~} n \to \infty.
    \eeq
Then the non-vanishing moments of the Gaussian $1$-matrix model are
    \beq
    \Gamma_{2n} = C_n \alpha^n \sim {(4\alpha)^n \over \sqrt{n^3 \pi}}
    {\rm ~~~as~~~} n \to \infty.
    \eeq

\subsubsection{Approximate solution by deforming the product}
\label{s-1-mat-gauss-deform-product}

In (\ref{e-gauss-1-mat-loop-eq}), $D_0$ is a derivation of $sh =
*_0$, not of $conc = *_1$. So it is not a differential equation. But
we can expand $*_1$ in a series in powers of $q(=1)$ around $*_0$.
Expanding $Z(\xi)$ also in a power series in $q$, turns the loop
equation into a sequence of differential equations in the shuffle
algebra. At order $q^0$, we get a nonlinear ODE for $Z_0(\xi)$.
Beyond that, we get a linear inhomogeneous ODE for $Z_k(\xi)$ in
terms of $Z_{k-1}(\xi)$. In the end, $q$ is set to $1$. Let us
illustrate this at ${\cal O}(q^0)$ and ${\cal O}(q^1)$. From section
\ref{s-1-mat-q-deformed-prod}, the expansion of $*_q$ around $*_0 =
sh$ is
    \beq
    (F *_q G)(\xi) = (F *_0 G)(\xi) - {q \over 2}~
    \xi *_0 \xi *_0 (D_0 F)(\xi) *_0 (D_0 G)(\xi) + \cdots.
    \eeq
Moreover $D_0 \xi = 1$, so keeping only terms to $O(q)$,
    \beq
    (Z *_q \xi) *_q Z &=& (Z *_0 \xi - {q \over 2}
        \xi *_0 D_0 Z *_0 \xi *_0 D_0 \xi)*_q Z \cr
    &=& Z *_0 \xi *_0 Z - {q \over 2} \xi *_0 \xi *_0 D_0 (Z *_0 \xi)
        *_0 D_0 Z - {q \over 2} \xi *_0 \xi *_0 D_0 Z *_0 Z \cr
    &=& \xi *_0 Z *_0 Z - {q \over 2} \bigg[ 2 \xi *_0 \xi *_0 Z *_0 D_0 Z
        + \xi *_0 \xi *_0 \xi *_0 D_0 Z *_0 D_0 Z \bigg].
    \eeq
So the LE are
    \beq
    D_0 Z &=& \alpha \bigg[\xi *_0 Z *_0 Z - {q \over 2}
        \bigg\{ 2 \xi *_0 \xi *_0 Z *_0 D_0 Z
        + \xi *_0 \xi *_0 \xi *_0 D_0 Z *_0 D_0 Z \bigg\} + {\cal
        O}(q^2)\bigg].
    \eeq
Suppose $Z(\xi) = Z_0(\xi) + q Z_1(\xi) + q^2 Z_2(\xi) + \ldots$.
Comparing coefficients of $q^0$ and $q^1$ we get
    \beq
    \ov{\alpha} D_0 Z_0 &=& \xi *_0 Z_0 *_0 Z_0 \cr
    \ov{\alpha} D_0 Z_1 &=& 2 \xi *_0 Z_0 *_0 Z_1
        - \half \bigg(2 \xi *_0 \xi *_0 Z_0 *_0 D_0 Z_0
         + \xi *_0 \xi *_0 \xi *_0 D_0 Z_0 *_0 D_0 Z_0 \bigg).
    \eeq
So we have a non-linear ODE for $Z_0(\xi)$, and linear
in-homogeneous ODEs for $Z_k, k \geq 1$. The boundary condition
$Z(0) = 1$ becomes $Z_0(0) = 1, ~Z_k(0) = 0, ~k \geq 1$.

{\flushleft \bf Zeroth order ${\cal O}(q^0)$. Replace concatenation
by shuffle product:} The ODE for $Z_0$ can be linearized by passing
to the shuffle reciprocal of $Z_0(\xi)$
    \beq
    Y(\xi) *_0 Z_0(\xi) = 1  &\Rightarrow&
    D_0 Z_0 = - Z_0 *_0 Z_0 *_0 D_0 Y.
    \eeq
$Y$ satisfies the inhomogeneous linear ODE $D_0 Y = - \a \xi$ with
boundary condition $Y(0) = 1$. So
    \beq
    Y(\xi) = 1 - {\a \over 2}~ \xi *_1 \xi = 1 - \a \xi^2.
    \eeq
Taking the shuffle reciprocal, we get (using $\xi^{*_0 n} = n! ~
\xi^{*_1 n} = n! ~ \xi^n$)
    \beq
    Z_0(\xi) = (1-{\a \over 2} \xi *_0 \xi)^{-1}
        = 1 + {\a \over 2} \xi *_0 \xi + ({\a \over 2})^2 \xi *_0 \xi *_0 \xi
        *_0 \xi + \cdots
    = \sum_{n=0}^\infty {\a^n \over 2^n} (2n)! ~\xi^{2n}.
    \eeq
So the generating function at order $q^0$ is
    \beq
    Z(\xi) = \sum_{n=0}^\infty \bigg({\alpha \over 2}
        \bigg)^n (2n)! ~\xi^{2n} + {\cal O}(q).
    \eeq
And the non-vanishing moments in this approximation are $G_{2n} =
\bigg( {\alpha \over 2} \bigg)^n (2n)! + {\cal O}(q)$. These are
compared with the exact moments in the table below.
    \beq
    \begin{tabular}{|c|c|c|}
  \hline
  Moments  & exact & ${\cal O}(q^0)$ \\
  \hline
  $G_2$ & $\alpha$ & $ \alpha$ \\
  $G_4$ & $2 \alpha^2$ & $6 \alpha^2$ \\
  $G_6$ & $5 \alpha^3$ & $90 \alpha^3$  \\
  $G_8$ & $14 \alpha^4$ & $2520 \alpha^4$ \\
  $G_{2n}, n \to \infty$ & ${(4 \alpha)^n \over \sqrt{\pi n^3}}$ & $({\alpha \over 2})^n (2n)!$  \\
  \hline
    \end{tabular}
    \eeq
Due to the $(2n)!$, the ${\cal O}(q^0)$ moments numerically exceed
the exact moments. We have a crude zeroth order answer with the
potential for calculating corrections. Of course, the gaussian is a
trivial model to solve. The value of our method lies in its
applicability to multi-matrix models for which no method of solution
exists.

\subsubsection{Approximate solution by deforming the left annihilation operator}
\label{s-1-mat-gauss-deform-derivative}

Next, we expand $D_0$ around $D_1$ so that the loop equation
(\ref{e-gauss-1-mat-loop-eq}) becomes a sequence of differential
equations with respect to $conc$. This leads to a different
approximation compared to section
\ref{s-1-mat-gauss-deform-product}, where we used the deformed
product. Here, the expansion parameter is $p=1-q$, which is
eventually set to $1$. Recall that the $q$-deformed annihilation
operator is
    \beq
    D_q F(x) = \sum_{k=1}^\infty {(- p \xi)^{k-1} \over k!} D_1^k F(\xi)
        = D_1 F(\xi) - {p \over 2} \xi D_1^2 F(\xi)  + {p^2 \over 6} \xi^2 D_1^3
            F(\xi) + {\cal O}(p^3).
    \eeq
If we expand $Z(\xi)$ in powers of $p$, $Z(\xi) = \sum_n Z_n(\xi)
p^n$, then
    \beq
    D_q Z(\xi) = \sum_{s=0}^\infty p^s \sum_{n=0}^s {(-1)^n \over
    (n+1)!} D_1^{n+1} Z_{s-n}(\xi) \xi^n
    \eeq
and
    \beq
    Z(\xi) *_1 \xi *_1 Z(\xi) = \sum_{s = 0}^\infty p^s
    \sum_{n=0}^s Z_n(\xi) *_1 \xi *_1 Z_{s-n}(\xi).
    \eeq
Comparing coefficients of $p$, we get a nonlinear ODE for $Z_0(\xi)$
and a sequence of $1^{\rm st}$-order linear ODEs for $Z_s(\xi)$ in
terms of the lower order ones $Z_{s-1}(\xi), \cdots$:
    \beq
    \sum_{n=0}^s {(-1)^n \over
    (n+1)!} D_1^{n+1} Z_{s-n}(\xi) \xi^n
    = \alpha \sum_{n=0}^s Z_n(\xi) *_1 \xi *_1 Z_{s-n}(\xi).
    \eeq
The first couple of orders are (all products are concatenation
products)
    \beq
    D_1 Z_0(\xi) &=& \alpha Z_0(\xi) \xi Z_0(\xi) \cr
    D_1 Z_1(\xi) - \half D_1^2 Z_0(\xi) \xi &=& 2 \alpha
        Z_0(\xi) \xi Z_1(\xi) \cr
    &\vdots&.
    \eeq
{\flushleft \bf Zeroth order:} At ${\cal O}(p^0)$ we have to solve
the ODE $D_1 Z_0(\xi) = \alpha ~ Z_0(\xi) *_1 \xi *_1 Z_0(\xi)$ with
$Z_0(0) = 1$. The solution is the $conc$ reciprocal
    \beq
    Z_0(\xi) = \ov{1- \half \alpha \xi *_1 \xi}
        = 1 + \frac{\alpha \,{\xi }^2}{2} +
  \frac{{\alpha }^2\,{\xi }^4}{4} +
  \frac{{\alpha }^3\,{\xi }^6}{8} +
  \frac{{\alpha }^4\,{\xi }^8}{16} +
  \frac{{\alpha }^5\,{\xi }^{10}}{32} + \cdots
        = \sum_{n=0}^\infty \bigg({\alpha \over 2} \bigg)^n
            \xi^{2n}.
    \eeq
The non-vanishing moments are thus
    \beq
    G_{2n} = \bigg({\alpha \over 2} \bigg)^n + {\cal O}(p).
    \eeq
These are compared with exact moments $\Gamma_{2n} = C_n \alpha^n
\sim {(4\alpha)^n \over \sqrt{n^3 \pi}}$, in Table
\ref{t-defor-derivative}. We see that at leading order, deforming
the annihilation operator underestimates the moments.

{\flushleft \bf Next to lowest order ${\cal O}(p^1)$:} At the next
order in $p=1-q$ we have an inhomogeneous linear first order ODE for
$Z_1(\xi)$
    \beq
    D_1 Z_1(\xi) - \half D_1^2 Z_0(\xi) \xi = 2 \alpha
        Z_0(\xi) \xi Z_1(\xi)
    \eeq
with boundary condition $Z_1(0) = 0$. Now $Y' + P Y + Q = 0$ has
solution
    \beq
    Y(\xi) = - I^{-1}(\xi) \int_0^\xi Q(\eta) I(\eta) d\eta
        ~ {\rm where~ } I(\xi) = \exp{\int_0^\xi P(\eta)}.
    \eeq
$Y = Z_1(\xi);~~ P = -2\alpha \xi Z_0(\xi);~~ Q = -\half \xi
Z_0''(\xi);~~ Z_0(\eta) = \ov{1 - \half \alpha \eta^2}; ~~ I(\xi) =
(1 - \half \alpha \xi^2)^2$. Thus,
    \beq
   Z_1(\xi) &=& - {3\alpha \xi^2 + 8 \log{(1-\half \alpha \xi^2)}
        \over 4 (1- \half \alpha \xi^2)^2}
        = \frac{\alpha \,{\xi }^2}{4} +
  \frac{{\alpha }^2\,{\xi }^4}{2} +
  \frac{25\,{\alpha }^3\,{\xi }^6}{48} +
  \frac{41\,{\alpha }^4\,{\xi }^8}{96} +
  \frac{99\,{\alpha }^5\,{\xi }^{10}}{320} + \cdots \cr
  &=& \ov{4} \alpha \xi^2 + \sum_{n \geq 2}
     \bigg({\alpha \xi^2 \over 2} \bigg)^n
     \bigg[{n \over 2} + 2 \sum_{r=0}^{n-2} ({r+1 \over n-r})\bigg].
    \eeq
To get the asymptotic behavior of moments for large $n$, let
$Z_1(\xi) = \sum \tilde G_{2n} \xi^{2n}$
    \beq
    \tl G_2 = {\alpha \over 4}, ~~~~
    \tl G_{2n} &=& \bigg({\alpha \over 2} \bigg)^n \bigg[{n \over 2}
        + 2 \sum_{r=0}^{n-2} ({r+1 \over n-r}) \bigg],~ n \geq 2
            \cr
    \Rightarrow ~~~ \tl G_{2n} &\sim& \bigg( {\alpha \over 2} \bigg)^n  \bigg[2n \log{n}
            - ({7 \over 2}-2\gamma) n + 2 \log{n} +{\cal O}(n^0)
            \bigg] , n \to \infty.
    \eeq
Recall that $Z(\xi) = Z_0(\xi) + p Z_1(\xi) + \cdots$ and $Z_0(\xi)
= \sum_n  ({\alpha \over 2})^n \xi^{2n}$. Combining, at ${\cal
O}(p)$ we have (after setting $p=1$)
    \beq
    G_2 &=& {3 \alpha \over 4} + {\cal O}(p^2);~~
    G_{2n} = \bigg({\alpha \over 2} \bigg)^n \bigg[1 + {n \over 2}
        + 2 \sum_{r=0}^{n-2} ({r+1 \over n-r}) \bigg] + {\cal O}(p^2),~ n \geq 2
            \cr
    G_{2n} &\sim& \bigg( {\alpha \over 2} \bigg)^n  \bigg[2n \log{n}
            - ({7 \over 2}-2\gamma) n + 2 \log{n} +{\cal O}(n^0)
            \bigg] + {\cal O}(p^2), n \to \infty.
    \eeq
This is to be compared with the exact moments
    \beq
    \Gamma_{2n} = {(2n)! \over n! (n+1)!} \alpha^n \sim {(4 \alpha)^n \over
    \sqrt{\pi n^3}}, ~~n \to \infty.
    \eeq
Going to the next to leading order in $p$ has improved the agreement
with the exact correlations. For large $n$, the next to leading
corrections to $G_{2n}$ are bigger in magnitude than the $0^{\rm
th}$ order $G_{2n}$. The accompanying table summarizes the
approximate correlations obtained by expanding the left annihilation
around the full annihilation operator in powers of $p=1-q$.
    \beq
    \begin{tabular}{|c|c|c|c|}
  \hline
  Moments  & exact & ${\cal O}(p^0)$ & ${\cal O}(p)$ \\
  \hline
  $G_2$ & $\alpha$ & $0.5 \alpha$ & $0.75 \alpha$ \\
  $G_4$ & $2 \alpha^2$ & $0.25 \alpha^2$ & $0.75\alpha^2$ \\
  $G_6$ & $5 \alpha^3$ & $0.125 \alpha^3$ & $0.646 \alpha^3$ \\
  $G_8$ & $14 \alpha^4$ & $0.0625 \alpha^4$ & $0.490 \alpha^4$ \\
  $G_{2n}, n \to \infty$ & ${(4 \alpha)^n \over \sqrt{\pi n^3}}$ & $({\alpha \over 2})^n$ & $({\alpha \over 2})^n (2n \log{n})$ \\
  \hline
    \end{tabular}
    \label{t-defor-derivative}
    \eeq

\subsection{Non-Gaussian $1$-matrix models} \label{s-1-mat-non-gaussian}

Recall that the $1$-matrix loop equation (\ref{e-1-mat-loop-eq}) for
a polynomial action $\tr S(A) = \tr \sum_{l=1}^m S_l A^l$ with $S_m
\ne 0$ determines higher rank correlations $G_{m-1}, G_m, G_{m+1},
\cdots$ in terms of the lower rank ones $G_0=1,G_1, G_2, \cdots
G_{m-2}$. Suppose we apply our approximation method here. At $0^{\rm
th}$ order we replace $conc$ by $sh$. Since left annihilation $D
\xi^n = \xi^{n-1}$ is a derivation of $sh$, the loop equation
becomes a quadratically non-linear ODE in the commutative shuffle
algebra
    \beq
    \sum_{l=1}^m l S_l D^{l-1} G(\xi) = G(\xi) \circ \xi \circ
    G(\xi).
    \eeq
However, for $m > 2$ (i.e. non-Gaussian models), the differential
operator $\sum_{l=1}^m l S_l D^{l-1}$ is {\em not} a derivation of
$sh$ and our trick of passing to the shuffle reciprocal does not
linearize this ODE. It can still be thought of as a set of recursion
relations (use $\xi^s *_0 \xi^t = {s+t \choose s} \xi^{s+t}$)
    \beq
    \sum_{l=1}^m l~ S_l~ G_{r + l -1} = \sum_{\stackrel{s+t+1 = r}{s,t \geq 0}}
    {r! \over s!~ t!}~ G_s ~G_t, ~~~~~{\rm for~~} r = 0,1,2,\ldots
    \eeq
which determine $G_{m-1}, G_m, G_{m+1}, \cdots$ in terms of $G_1,
G_2, \cdots G_{m-2}$:
    \beq
    r=0: && S_1 G_0 + 2 S_2 G_1 + \cdots + m S_m G_{m-1} =0 \cr
    r=1: && S_1 G_1 + S_2 G_2 + \cdots + S_m G_m = 1, {\rm
    ~~~e.t.c.}
    \eeq
Our approach {\em does not} lead to a significant simplification for
non-Gaussian $1$-matrix models. However, we observe that the passage
to the limit $q=0$ (replacement of $conc$ by $sh$) did not change
the dimension of the space of solutions to the original loop
equations.

\section{Approximation method for multi-matrix models}
\label{s-approx-meth-multi-mat}

Recall the multi-matrix LE (\ref{e-loop-eqns-in-terms-of-left-ann})
for the generating series of gluon correlations ${\cal S}^i G(\xi) =
G(\xi) \xi^i G(\xi)$ where ${\cal S}^i = \sum_{n \geq 0} (n+1)
S^{j_1 \cdots j_n i} D_{j_n} \cdots D_{j_1}$. Products on the RHS
are $conc$ products, but $D_j$ are not derivations of $conc$. So the
LE are not differential equations. By analogy with $1$-matrix
models, two ways around this mismatch come to mind. We could
$p$-expand $D_i$ around full annihilation $\bfD_i$, which {\em is} a
derivation of $conc$. Or, we could $q$-expand $conc$ around $sh$,
with respect to which $D_i$ is a derivation. Both these turn LE into
quadratically non-linear PDEs at $0^{\rm th}$ order in $p$ or $q$.
In the former approach these are PDEs on the non-commutative
concatenation algebra, while in the latter case, they are PDEs on
the commutative shuffle algebra. We focus on the second approach in
section \ref{s-multi-mat-zeroth-ord-q} due to its similarity with
deformation quantization, and briefly consider the first approach in
section \ref{s-multi-mat-zeroth-ord-p}. Beginnings of a formalism to
go beyond zeroth order are in section
\ref{s-multi-mat-beyond-zeroth-order}.


\subsection{Multi-matrix LE at ${\cal O}(q^0)$ and the shuffle reciprocal}
\label{s-multi-mat-zeroth-ord-q}

At $0^{\rm th}$ order in $q$, we replace $conc$ by $sh$. Then the
factorized LE (\ref{e-loop-eqns-in-terms-of-left-ann})
become\footnote{Here, we use $\circ$ for $*_0 = sh$ to avoid
subscripts.}
    \beq
    {\cal S}^i G(\xi) = G(\xi) \circ \xi^i \circ G(\xi)
    \label{e-zeroth-order-loop-eq}
    \eeq
with the boundary condition $G(0) = 1$.
(\ref{e-zeroth-order-loop-eq}) is a quadratically non-linear PDE on
the shuffle algebra. In general, the order of the PDE is one less
than the degree of $S(A)$. If ${\cal S}^i$ is a derivation of $sh$,
we can change variables so that (\ref{e-zeroth-order-loop-eq})
becomes a linear PDE for the shuffle reciprocal of $G(\xi)$ denoted
$F(\xi)$, $F(\xi) \circ G(\xi) = 1$. The shuffle reciprocal exists
as a formal series since the constant term $G_0 = 1$ does not
vanish. Moreover, since $G_I$ are cyclic and shuffle product
preserves cyclicity, $F_I$ are also cyclic. Assuming $S^i$ is a
derivation of $sh$,
    \beq
    {\cal S}^i (F(\xi) \circ G(\xi))) =0 ~~\Rightarrow~~
    F \circ {\cal S}^i G = - {\cal S}^i F \circ G
    ~~\Rightarrow~~ {\cal S}^i G &=& - G \circ {\cal S}^i F \circ
    G.
    \eeq
Putting this in (\ref{e-zeroth-order-loop-eq}) we get $G \circ {\cal
S}^i F \circ G = G \circ \xi^i \circ G$. Shuffle multiplying by $F
\circ F$ reduces the LE to a system of inhomogeneous linear PDEs in
the shuffle algebra
    \beq
        {\cal S}^i F(\xi) = - \xi^i.
    \label{e-sh-reciprocal-loop-eq}
    \eeq
We call these {\em shuffle reciprocal LE}. We seek cyclically
symmetric solutions to them. The LHS of
(\ref{e-sh-reciprocal-loop-eq}) is the same as in the LE
(\ref{e-loop-eqns-in-terms-of-left-ann}) with $G$ replaced by its
reciprocal $F$. The $-$ sign due to inversion has been written on
the RHS. The RHS, however is much simpler than in
(\ref{e-loop-eqns-in-terms-of-left-ann}) since the quadratic factor
in $G(\xi)$ has been eliminated. For the zero-momentum Gaussian,
Chern-Simons and Yang-Mills matrix models we get (from section
\ref{s-derivation-property})
    \beq
    Gaussian && C^{ij}~ D_j F(\xi) ~=~ -\xi^i \cr
    Chern-Simons && i \k ~\eps^{ijk} [D_k,D_j] F(\xi) ~=~ -\xi^i \cr
    Yang-Mills && - \ov{\a}~g^{ik} g^{jl} [D_j,[D_k,D_l]] F(\xi) ~=~
    -\xi^i.
    \label{e-linear-loop-eq-g-cs-ym}
    \eeq
Thus, we have used the derivation properties of these theories to
effectively linearize the LE at order $q^0$. We still have to solve
these linear PDEs on the $\infty$-dimensional vector space spanned
by $\xi^I$. First we find a formula to recover $G(\xi)$ from its
shuffle reciprocal $F(\xi)$.
    \beq
    (F \circ G)(\xi) = 1 \Rightarrow \sum_{I = J \sqcup K} F_J G_K =
    \delta^I_\emptyset.
    \eeq
We can solve these equations starting from $G_0 = F_0 =1$. The first
few equations are
    \beq
    F_i + G_i = 0, && F_{ij} + F_i G_j  + F_j G_i + G_{ij} = 0, \cr
    F_{ijk} + F_{ij} G_k + F_{ik} G_j &+& F_{jk} G_i + F_i G_{jk} +
    F_j G_{ik} + F_k G_{ij} + G_{ijk} = 0, ~~~ \ldots
    \eeq
Since each successive equation involves the next higher rank $G_{I}$
only linearly, we need only solve a linear equation at each step.
Thus for $|I| > 0$,
    \beq
    G_{I} = - \sum_{I = J \sqcup K, K \ne I} F_J G_K
    \eeq
expresses higher rank $G_I$ in terms of lower rank ones and the
reciprocal $F$. Iterating,
    \beq
    G_I = \sum_{n=1}^{|I|} (-1)^n \sum_{\stackrel{I = I_1 \sqcup I_2 \sqcup \cdots \sqcup
    I_n}{I_k ~\ne~ \emptyset~ \forall~ k}} F_{I_1} F_{I_2} \cdots
    F_{I_n} ~~~ {\rm for~~} I~ \ne~ \emptyset.
    \label{e-shuffle-reciprocal}
    \eeq
$I = I_1 \sqcup I_2 \sqcup \cdots \sqcup I_n$ $\Leftrightarrow$
$I_1, \cdots, I_n$ are complementary order-preserving subwords of
$I$. For example, $G_i = - F_i$,  $    G_{ij} = - F_{ij} + 2 F_i
F_j$ and
    \beq
    G_{ijk} &=&  - F_{ijk} + 2(F_i F_{jk} + F_j F_{ik} + F_k F_{ij}) - 6 F_i F_j F_k
        \cr
    G_{ijkl} &=& -F_{ijkl} + 2 (F_i F_{jkl} + F_j F_{ikl} + F_k F_{ijl} + F_l F_{ijk}
        + F_{ij} F_{kl} + F_{ik} F_{jl} + F_{il} F_{jk}) \cr &&
        - 6(F_i F_{jk} F_l + F_j F_{ik} F_l
        F_i F_{jl} F_k + F_j F_{il} F_k + F_k F_{ij} F_l + F_i F_{kl}
        F_j) + 24 F_i F_j F_k F_l.
    \eeq
This formula shows that the mapping to shuffle reciprocal (for
series with non-vanishing constant term) is one-to-one. We don't
lose any information in going from $G(\xi)$ to $F(\xi)$ and back.
Once we solve (\ref{e-sh-reciprocal-loop-eq}), for $F(\xi)$ we may
straightforwardly recover $G_I$ using (\ref{e-shuffle-reciprocal}).

\subsubsection{Solution of Gaussian multi-matrix model at zeroth order in $q$}
\label{s-gauss-multi-mat-zeroth-order-in-q}

Consider the Gaussian multi-matrix model $\tr S(A) = \half \tr
C^{ij} A_i A_j$ with symmetric covariance $C^{ij} = C^{ji}$. At
$0^{\rm th}$ order in $q$, the shuffle reciprocal LE
(\ref{e-linear-loop-eq-g-cs-ym}) are
    \beq
    D_k F(\xi) = -C_{kj} \xi^j,
    \label{e-gauss-zeroth-ord-linear-loop-eq}
    \eeq
where $C_{kj} = C_{jk}$ is the matrix inverse of $C^{ij}$. We seek a
solution of (\ref{e-gauss-zeroth-ord-linear-loop-eq}) of the general
form
    \beq
    F(\xi) = 1 + F_{i_1} \xi^{i_1} + F_{i_1 i_2} \xi^{i_1 i_2} +
    \cdots + F_{i_1 \cdots i_n } \xi^{i_1 \cdots i_n} + \cdots,
    \eeq
where $F_I$ are cyclically symmetric. $G_0 =1$ fixes $F_0 = 1$.
Substituting in (\ref{e-gauss-zeroth-ord-linear-loop-eq}) using $D_i
\xi^{i_i \cdots i_n} = \d^{i_1}_i \xi^{i_2 \cdots i_n}$ we get
    \beq
    F_i + F_{i i_2} \xi^{i_2} + F_{i i_2 i_3} \xi^{i_2 i_3} + \cdots
    +  F_{i i_2 \cdots i_n} \xi^{i_2 \cdots i_n} + \cdots = C_{ij}
        \xi^j.
    \eeq
Comparing coefficients of words $\xi^I$ we read off the solution
    \beq
    F_i = 0, ~~~ F_{ij} = -C_{ij}, ~~~ F_{i_1 \cdots i_n} = 0 {\rm
    ~~for~ ~} n \geq 3.
    \eeq
The solution is a quadratic polynomial $F(\xi) = 1 - C_{ij}
\xi^{ij}$. Using (\ref{e-shuffle-reciprocal}) we get
    \beq
    G_0 = 1,~~ G_i = 0,~~ G_{ij} = - F_{ij} = C_{ij},~~ G_{ijk} =
    0,
    G_{ijkl} = 2 \{ C_{ij} C_{kl} + C_{ik} C_{jl} + C_{il} C_{jk} \},
    \ldots
    \eeq
Thus, for the Gaussian multi-matrix model, the linear equations
(\ref{e-gauss-zeroth-ord-linear-loop-eq}) for shuffle reciprocal,
along with the boundary condition $F_0 = 1$ have a unique solution.
Comparing with exact moments from the planar Wick theorem, $\G_0 =
1,~ \G_{ij} = C_{ij},~ \G_{ijkl} = C_{ij} C_{kl} + C_{il} C_{jk},
\cdots$, we see that the approximation is an over estimate (as we
found in the $1$-matrix example in section
\ref{s-1-mat-gauss-deform-product}).

\subsubsection{Chern-Simons matrix model at zeroth order in $q$}
\label{s-cs-zeroth-ord-q}

Consider the zero-momentum limit of $3$d Chern-Simons(CS) gauge
theory. This corresponds to the $3$-matrix model with action $\tr
S(A) = 2 i \k \tr A_1 [A_2, A_3]$. Such an action also results from
considering terms in (\ref{e-gauge-fixed-ym-action}) that are linear
in momentum. The $0^{\rm th}$ order CS loop equation
(\ref{e-linear-loop-eq-g-cs-ym}) for the shuffle reciprocal $F(\xi)$
is
    \beq
    i \k ~ \eps^{ijk} [D_k,D_j] F(\xi) = -\xi^i {\rm ~~~~ or ~~~~}
    2 i \k ~\eps^{ijk} D_k D_j F(\xi) = -\xi^i.
    \label{e-linear-loop-eq-cs}
    \eeq
We seek a solution to (\ref{e-linear-loop-eq-cs}) among formal
series $F(\xi) = F_I \xi^I$ with cyclic coefficients $F_I$
satisfying $F_I^* = F_{\bar I}$. Eqn.(\ref{e-linear-loop-eq-cs}) is
an inhomogeneous $2^{\rm nd}$ order linear PDE in an infinite
dimensional space spanned by the words $\xi^I$. $F_0 =1$ does not
suffice to fix a solution. For example, $F_i$ are undetermined,
since $\xi^i$ is annihilated by the LHS. Inserting $F_I \xi^I$ into
(\ref{e-linear-loop-eq-cs}) gives
    \beq
    2 i \k ~\eps^{i i_1 i_2} F_{i_1 \cdots i_n} \xi^{i_3 \cdots i_n} =
    - \xi^i.
    \eeq
The PDEs become linear equations for the coefficients $F_I$ with
$|I| \geq 2$,
    \beq
    n =2:  ~~~ \eps^{ijk} F_{jk} = 0; ~~~
    n =3:  ~~~ 2 i \k~ \eps^{ijk} F_{jkl} = - \d^i_l; ~~~{\rm and}~~~
    n > 3: ~~~ \eps^{i i_1 i_2} F_{i_1 \cdots i_n} = 0.
    \label{e-CS-inhom-linear-eqns}
    \eeq
Being a system of inhomogeneous linear equations, the general
solution is the sum of a particular solution and the general
solution of the corresponding homogeneous system. A particular
solution with minimal number of non-vanishing $F_I$ is
    \beq
    F_0 = 1, ~~~ F_{123} = F_{231} = F_{312} =
    F_{321}^* = F_{213}^* = F_{132}^*
    = {i \over 4 \k} {\rm ~~~and~~}
    F_I = 0 ~~ \forall {\rm ~~other~~} I.
    \eeq
To see this we need only consider $n=3$, where despite appearances,
after accounting for cyclic symmetry, there are only a pair of
independent equations, the real and imaginary parts of
    \beq
    F_{321} - F_{123} = \ov{2 i \k}.
    \eeq
By hermiticity, $F^*_{321} = F_{123}$ or $\Re F_{321} = \Re F_{123}$
and $\Im F_{321} = - \Im F_{123}$. Since $\k$ is real, the real part
of the above equation is an identity, so the real part $\Re F_{123}
= \Re F_{321}$ is left undetermined, and we can set it to zero. Its
imaginary part gives $\Im F_{123} = \ov{4 \k}$, which is the
advertised particular solution. For this particular solution the
gluon green functions at order $q^0$ can be non-trivial only if
their rank is divisible by $3$. For example, $G_0 =1$,
    \beq
    G_i &=& -F_i = 0,~ G_{ij} =0, ~~
    G_{123}=G_{231}=G_{312} = G_{321}^* = G_{132}^* = G^*_{213} = \ov{4 i \k}, \cr
    G_{ijk} &=& 0 {\rm ~otherwise~},~ G_{113322} = -8 G_{132} F_{132} = -\ov{2\k^2},
    ~~~ G_{112233} =0, {\rm ~~ etc.}
    \eeq
Let us now consider the general solution to the inhomogeneous linear
equations (\ref{e-CS-inhom-linear-eqns}). It is straightforward to
see that they have infinitely many solutions, since the
corresponding homogeneous equations $\eps^{i i_1 i_2} F_{i_1 i_2
\cdots i_n} = 0,~~~ n \geq 2$ do. Indeed, any tensor $F_{i_1 i_2 i_3
\cdots i_n}$ that is symmetric under interchange of a pair of
adjacent indices is a solution to the homogeneous equations. By
cyclic symmetry, the two indices can be chosen as $i_1$ and $i_2$.
Then such an $F_{i_1 i_2 i_3 \cdots i_n}$ is annihilated due to
antisymmetry of $\eps^{i i_1 i_2}$. Even after imposing hermiticity
and cyclic symmetry, this will leave an infinite number of
homogeneous solutions, for example any totally symmetric real tensor
$F_I$ is automatically cyclically symmetric and satisfies $F_I^* =
F_{\bar I}$. To get an idea of how many solutions there are among
tensors of a fixed rank, consider each rank individually since the
equations do not mix tensors of different rank. For $n=1$, we do not
have any LE, but hermiticity implies that $F_1, F_2, F_3$ are three
arbitrary real quantities. For $n=2$, $\eps^{ijk} F_{jk} = 0$ does
not impose any condition on $F_{11}, F_{22}, F_{33}$, which are real
by hermiticity, and says that $F_{12}, F_{23}$ and $F_{31}$ are
symmetric tensors, which must again be real. For $n=3$, as we saw
earlier, $2 i \k ~\eps^{ijk} F_{jkl} = -\d^i_l$ is just the single
condition $\Im F_{123} = \ov{4 \k}$. After accounting for cyclicity
and hermiticity, there are $11$ independent components of $F_{ijk}$.
The $10$ undetermined components can be taken as the real numbers
    \beq
    F_{111}, F_{222}, F_{333}, \Re F_{123},
    F_{122}, F_{233}, F_{311}, F_{133}, F_{211}, F_{322}.
    \eeq
For $n=4$, accounting for cyclic $F_I$, there are only $9$
conditions
    \beq
    F_{1123} = F_{1132} = F_{1213}, ~~
    F_{2231} = F_{2213} = F_{2321}, &&
    F_{3312} = F_{3321} = F_{3132}, \cr
    F_{1212} = F_{1122}, ~~ F_{2323} = F_{2233},
    && F_{3131} = F_{3311}.
    \eeq
But there are $c(n=4,\La=3) = 24$ independent cyclic symmetric
fourth rank tensors (see appendix \ref{a-cyclic-tensors}). Thus we
have a large space of homogeneous solutions among fourth rank
tensors. A similar situation continues for $n > 4$. The LE at order
$q^0$ (\ref{e-linear-loop-eq-cs}), though linear and easy to solve,
have infinitely many solutions. As explained in section
\ref{s-non-uniq-of-loop-eq-solns}, this is true of the original LE
and is not an artifact of our approximation scheme. It remains to
see if the additional equations obtained in \ref{s-additional-eqns}
fix this shortcoming.

\subsubsection{Yang-Mills multi-matrix model at zeroth order in $q$}
\label{s-ym-zeroth-ord-q}

Consider the Yang-Mills matrix model with action $\tr S(A) =
-\ov{4\a} \tr [A_i,A_j][A_k,A_l] g^{ik} g^{jl}$. The LE for the
shuffle reciprocal $F(\xi)$ of the moment generating series $G(\xi)$
at zeroth order in $q$ are
    \beq
    g^{ik} g^{jl} [D_j,[D_k,D_l]] F(\xi) = \a ~ \xi^i ~~~{\rm for}~~~ 1
        \leq i \leq \La.
    \label{e-linear-loop-eq-ym}
    \eeq
Interesting special cases are $\La =4,2$ which correspond to the
zero momentum limit of $4$ and $2$ dimensional large-$N$ Yang-Mills
theory. For $\La = 2$ and a flat Euclidean metric $g^{ij} =
\d^{ij}$, the matrix model action is $\tr S(A) = - \ov{2 \a} \tr
[A_1,A_2]^2$. (\ref{e-linear-loop-eq-ym}) are a system of $\La$
inhomogeneous $3^{\rm rd}$ order linear PDEs for $F(\xi) = F^I
\xi_I$ which is normalized to $F_0 = 1$. $F_I$ must be cyclically
symmetric. We need additional conditions to fix a solution since any
quadratic polynomial is annihilated by the LHS, so $F_i$ and
$F_{ij}$ are not fixed by (\ref{e-linear-loop-eq-ym}). Let us assume
the metric $g^{ij} = \d^{ij}$ and not make a distinction between
lower and upper indices, with repeated indices being summed. Then
(\ref{e-linear-loop-eq-ym}) becomes (using the short-hand $D_{ijk}=
D_i D_j D_k$)
    \beq
    (2 D_{jij} - D_{jji} - D_{ijj}) F(\xi) = \a~ \xi_i
    &\Rightarrow& (2 F_{jij i_4 \cdots i_n} - F_{ijj i_4 \cdots i_n} - F_{jji i_4 \cdots i_n})
    \xi^{i_4 \cdots i_n} = \a ~\xi_i.
    \eeq
Comparing coefficients we get these conditions
    \beq
    n = 3  &\Rightarrow& 2 F_{jij} - F_{ijj} - F_{jji} = 0  ~~\forall~~
    i \cr
    n = 4 &\Rightarrow& 2 F_{jijk} - F_{ijjk} - F_{jjik} =  \a~ \d_{ik}
        ~~\forall~~ i,k \cr
    n \geq 5 &\Rightarrow& 2 F_{jij i_4 \cdots i_n} - F_{ijj i_4 \cdots i_n}
        - F_{jji i_4 \cdots i_n} = 0  ~~\forall~~ i, i_4 \cdots i_n.
    \eeq
The condition for $n=3$ is an identity for cyclically symmetric
tensors, so we drop it. These are infinitely many linear equations
for the tensors $F_I$. A major simplification is that the equations
do not mix tensors of different ranks, i.e. the matrix defining the
system is block diagonal with all blocks finite dimensional. Let us
specialize to the simplest non-trivial case of the $\La =2$ matrix
model. We will show that a particular (cyclically symmetric)
solution is
    \beq
    F_0 = 1,~~ F_{1122} = F_{2112} = F_{2211} = F_{1221}
        = -{\a \over 2} {\rm ~~and~the~remaining} ~~ F_I = 0.
    \eeq
The only non-trivial part of this particular solution involves the
rank $n=4$ tensors. The equations for the rest are homogeneous and
they can be set to zero. For $n = 4$ we need to find a solution to
$2 F_{jijk} - F_{ijjk} - F_{jjik} =  \a~ \d_{ik}$. These look like
four equations,
    \beq
        2 F_{2121} - F_{1221} - F_{2211} = \a, &&
        2 F_{2122} - F_{1222} - F_{2212} = 0, \cr
        2 F_{1212} - F_{2112} - F_{1122} = \a, &&
        2 F_{1211} - F_{2111} - F_{1121} = 0.
    \eeq
But there is only one independent non-trivial condition after
accounting for cyclic symmetry
    \beq
    F_{1122} - F_{1212} = -{\a \over 2}.
    \eeq
Thus we see that $F_0=1, F_{1122}$ and cyclic permutations $=-\a/2$
and all other $F_I =0$ is a particular solution. The gluon green
functions at order $q^0$ are obtained via the shuffle reciprocal
(\ref{e-shuffle-reciprocal}) which imply that non-vanishing
correlations have rank divisible by $4$, for example,
    \beq
    G_0 = 1,~~ G_i = G_{ij} = G_{ijk} =0,~~ G_{1122} {\rm ~and ~cyclic~} = {\a \over 2},
    {\rm ~and~ other~} G_{ijkl} =0,~ {\rm ~~etc}.
    \eeq
Now comes the harder question of the general solution of the
homogeneous linear system \footnote{Recall that $n=3$ was
identically satisfied by cyclically symmetric tensors.}
    \beq
    n = 4 &\Rightarrow& 2 F_{jijk} - F_{ijjk} - F_{jjik} =  0
        ~~\forall~~ i,k \cr
    n \geq 5 &\Rightarrow& 2 F_{jij i_4 \cdots i_n} - F_{ijj i_4 \cdots i_n}
        - F_{jji i_4 \cdots i_n} = 0  ~~\forall~~ i, i_4 \cdots i_n
    \eeq
For $n=4$, as we saw before, there is only one non-trivial equation
$F_{1122} = F_{1212}$. But there are $c(n=4,\La=2)= 6$ independent
cyclically symmetric rank $4$ tensors (see appendix
\ref{a-cyclic-tensors}) which can be taken as $F_{2222},~~
F_{1222},~~ F_{1122},~~ F_{1212},~~ F_{1112},~~ F_{1111}$.
Hermiticity $F_I = F^*_{\bar I}$ implies that all of them are real
since reversal of order of indices can be achieved by cyclic
permutations in each case. Thus the general solution for rank four
tensors assigns 5 arbitrary real parameters to $F_{2222},~~
F_{1222},~~ F_{1122}= F_{1212},~~ F_{1112}$ and $F_{1111}$.

For $n=5$, once the dust settles, there are only two non-trivial
equations
    \beq
    F_{11122} = F_{11212} {\rm ~~~and~~~} F_{11222} = F_{12122}
    \eeq
Taking account of cyclic symmetry, there are $c(5,2)=8$ independent
rank $5$ tensors, which can be taken as $F_{22222}, F_{12222},
F_{11222}, F_{12122}, F_{11122}, F_{11212}, F_{11112}$ and
$F_{11111}$. In general these are complex, but hermiticity and
cyclicity imply they are all real. Thus we have two linear
constraints on $8$ real parameters and therefore a six
real-dimensional space of solutions to the shuffle reciprocal LE for
rank $5$ tensors:
    \beq
    F_{22222},~~ F_{12222},~~
    F_{11222} = F_{12122},~~ F_{11122}= F_{11212},~~ F_{11112},~~ F_{11111}
    \eeq
are freely specifiable real quantities.

This abundance of solutions continues to hold for $n \geq 6$. It is
easy to see that the homogeneous linear equations $2 F_{jijI} -
F_{ijjI} - F_{jjiI} =0$ have an infinite number of solutions.
Observe that any tensor that is totally symmetric in any three
adjacent indices\footnote{By cyclic symmetry, those three indices
can be taken as the first three.} satisfies this equation. In
particular, totally symmetric tensors are an infinite class of
solutions. The underdetermined nature of the linear equations for
$F(\xi)$ is not an artifact of our approximation scheme. It is
already true of the full LE as shown in section
\ref{s-non-uniq-of-loop-eq-solns}. It remains to implement the
additional conditions (\ref{e-aditional-eq}) to see if they select a
solution.

\subsection{LE with deformed left annihilator and
the concatenation reciprocal}
\label{s-multi-mat-zeroth-ord-p}

We also have the option of approximating the LE
(\ref{e-loop-eqns-in-terms-of-left-ann}) by replacing left
annihilation $D_i$ by full annihilation $\bfD_i$ at zeroth order in
an expansion in powers of $p=1-q$. Since $\bfD_i$ is a derivation of
$conc$, this again turns the LE into non-linear PDEs, but this time
on the non-commutative free algebra. As before, it is possible to
convert the non-linear PDEs into linear PDEs by passage to the
concatenation reciprocal. Recall that ${\cal S}^i = \sum_{n} (n+1)
S^{j_1 \cdots j_n i} D_{j_n} \cdots D_{j_1}$. When $D_j$ is replaced
by $\bfD_j$, we denote the resulting differential operator
    \beq
    \bfS^i = \sum_{n} (n+1) S^{j_1 \cdots j_n i}
        \bfD_{j_n} \cdots \bfD_{j_1}.
    \eeq
Moreover, assume couplings $S^I$ are such that $\bfS^i$ is a linear
combination of iterated commutators of $\bfD_j$ and therefore a
derivation of $conc$. This is the case for the Gaussian, CS and YM
matrix models or any linear combination thereof. At zeroth order in
$p$, the LE become
    \beq
    \bfS^i G(\xi) = G(\xi) \xi^i G(\xi).
    \label{e-zeroth-order-concat-loop-eqn}
    \eeq
Now we'd like to use the same trick as before and turn this into a
linear equation for the $conc$ reciprocal of $G(\xi)$. Though $conc$
is non-commutative, left and right concatenation reciprocals of
$G(\xi)$ are both equal. Let $GR = 1$ and $LG =1$. Multiplying the
first equation by $L$ from the left and using the second, we get
$R=L$. So let $F(\xi)$ be the unique two-sided $conc$
reciprocal\footnote{$F_0 = G_0=1$} of $G(\xi)$. Assuming $\bfS^i$ is
a derivation of $conc$, $\bfS^i(FG) = (\bfS^i F)G + F (\bfS^i G)
=0$. This turns (\ref{e-zeroth-order-concat-loop-eqn}) into a linear
equation for the $conc$ reciprocal
    \beq
    \bfS^i F(\xi) = - \xi^i.
    \label{e-concat-reciprocal-loop-eqn}
    \eeq
Inserting $F(\xi) = F_I \xi^I$ into
(\ref{e-concat-reciprocal-loop-eqn}) gives linear equations for
coefficients $F_I$. Once $F_I$ are determined, we recover $G_I$ at
zeroth order in $p$ using the following formula for $conc$
reciprocal.
    \beq
    (FG)_I = \d_{I,\emptyset} ~\Rightarrow~
    G_0 = 1 {\rm ~~and~~} \d_I^{I_1 I_2} F_{I_1} G_{I_2} = 0
    ~\Rightarrow~ G_I = -\sum_{\stackrel{I = I_1 I_2,}{ I_1 \ne \emptyset}} F_{I_1} G_{I_2}
    {\rm ~~for~~} |I| > 0.
    \eeq
Iterating this, we solve for $G_I$
    \beq
    G_I = \sum_{n=1}^{|I|} (-1)^n \sum_{\stackrel{I=I_1 I_2 \cdots I_n}
    {I_k ~\ne~ \emptyset~~ \forall k}} F_{I_1} F_{I_2} \cdots F_{I_n}
    {\rm ~~~ for ~~~} I \ne \emptyset.
    \eeq
For example, the first few gluon correlations are
    \beq
    G_0 = 1;~~ G_i = -F_i;~~ G_{ij} = -F_{ij} + F_i F_j;~~
    G_{ijk} = -F_{ijk} + F_{ij} F_k + F_i F_{jk} - F_i F_j F_k;
    ~ \ldots
    \eeq
Thus $conc$ reciprocal is a $1-1$ map. However, unlike shuffle
reciprocal, it does not preserve cyclicity. Though we do not lose
any information in the passage from $G$ to $F$, the cyclic property
of $G_I$ gets slightly garbled when expressed in terms of $F_I$. For
example, cyclic symmetry of $G_{ijk}$ implies the relation $F_{ij}
F_k - F_{ijk} = F_j F_{ki} - F_{jki}$. Thus, we should look for
solutions to (\ref{e-concat-reciprocal-loop-eqn}) among $F_{I}$ that
lead to cyclically symmetric $G_J$'s. This makes identifying the
appropriate solutions of (\ref{e-concat-reciprocal-loop-eqn})
potentially harder than for the corresponding shuffle reciprocal LE
(\ref{e-sh-reciprocal-loop-eq}). There is another reason why the
concatenation reciprocal LE (\ref{e-concat-reciprocal-loop-eqn}) are
a potentially harder infinite linear system to solve than their
shuffle reciprocal counter part (\ref{e-zeroth-order-loop-eq}). Left
annihilation acting on a monomial produces a monomial $[D_j F]_I =
F_{jI}$. But due to its democratic nature, full annihilation
produces a linear combination of monomials $[\bfD_j F]_I = \d_I^{I_1
I_2} F_{I_1 j I_2}$. Thus the matrix defining the system of linear
equations for $F_I$ would be less sparse than before. Nevertheless,
the moral is that replacing $D_j$ by $\bfD_j$ at $0^{\rm th}$ order
in an expansion in $p$ allows for an effective linearization of the
LE provided the action has the derivation property.


\subsection{Formalism for multi-matrix models beyond zeroth order}
\label{s-multi-mat-beyond-zeroth-order}

At ${\cal O}(q^0)$ our approximation amounted to replacement of
non-commutative $conc$ by commutative $sh$ in the LE. This is like
approximating the associative product of operators in quantum
mechanics by a commutative product of functions on phase space. To
go beyond this, we need a formula expressing $conc$ as a series
around $sh$, by analogy with the Moyal $*$-product formula
    \beq
    (\tl F *_\hbar \tl G)(x,p) &=& \sum_{n=0}^\infty
        \bigg({-i \hbar \over 2} \bigg)^n \ov{n \!} \{ \tl F, \tl
        G\}_{(n)} ~=~ \tl F \tl G - {i \hbar \over 2}\{\tl F,\tl G\} + \cdots,
         ~~~~ {\rm where } \cr
    \{ \tl F, \tl G\}_{(n)} &=& \sum_{r=0}^n (-1)^r \tl F^{j_1 \cdots
    j_r}_{i_1 \cdots i_{n-r}} \tl G^{i_1 \cdots i_{n-r}}_{j_1 \cdots j_r}
    {\rm ~~ with ~~} \tl F^i = \dd{\tl F}{p_i}, ~~ \tl F_i = \dd{\tl A}{x^i}, {\rm
    ~~~etc~~}
    \eeq
for the symbols of operators (here Weyl ordered) in quantum
mechanics. The first non-trivial term in such a formula involves the
classical Poisson bracket. So one strategy is to look for a natural
Poisson bracket on the shuffle algebra. However, there are
differences from the usual situation where Heisenberg equations are
approximated by Hamilton's equations. While the Heisenberg equations
of quantum mechanics involve commutators of the associative product,
the LE directly involve the associative concatenation product and
not its commutator. Another difference from the usual situation in
deformation quantization is that we know the product at both $q=0$
and $q=1$ whereas one usually knows the product only at $\hbar =0$.
Once we have such a formula, then as we did for $1$-matrix models
(section \ref{s-1-mat-gauss-deform-product}), we would expand the
generating series of gluon correlations $G(\xi)= \sum_{k=0}^\infty
G^{(k)}(\xi) q^k$ in a power series in $q$ and find equations for
the $G^{(k)}(\xi)$ order by order in $q$, starting from the $0^{\rm
th}$ order equations for $G^{(0)}(\xi)$ of section
\ref{s-multi-mat-zeroth-ord-q}. However, the situation for
multi-matrix models is substantially more complicated than for the
$1$-matrix models of section \ref{s-approx-meth-one-mat}. This is
because $conc$ is non-commutative while it was commutative in the
single-matrix case.

\subsubsection{$q$-Deformed product and Poisson bracket on shuffle algebra}
\label{s-q-prod-and-poisson-bracket}

We exhibit a $1$-parameter family of associative products $*_q$ that
interpolate between commutative shuffle $*_0$ and concatenation
$*_1$. It reduces to the $q$-product for a single generator
introduced in (\ref{e-def-q-prod-one-mat}) and is defined as $(F *_q
G)(\xi) = [F *_q G]_I \xi^I$ where\footnote{To avoid too much
clutter we will occasionally drop the subscript in $*_q$ and
indicate it by $*$.}
 \beq
    [F *_q G]_I \equiv \sum_{J \sqcup K = I} p^{\chi(I,J,K)}
    F_J G_K ~~~~~~ {\rm and~~~~} p = 1-q.
 \eeq
The (two-word) crossing number $\chi(I;J,K)$ of the ordered triple
$\{I; J, K\}$ is the {\it minimum} number of transpositions of
elements of $J$ and $K$ in order to transform $JK$ into $I$ when $J$
and $K$ are order-preserving sub-words of $I$. For example,
    \beq
    \chi(ijk;i,jk) = 0, ~~~ \chi(ijk;ik,j) = 1,
        ~~~ \chi(ijk;jk,i) = 2.
    \eeq
For $q=1~(p=0)$, this formula reduces to $conc$. For, the only term
that contributes is the one with $\chi(I;J,K)=0$ i.e. no crossings,
so $I=JK$. Then
    \beq
    (F *_1 G)_I = \d_I^{JK} F_J G_K.
    \eeq
If $q=0~(p=1)$, then $p^{\chi(I;J,K)} = 1$ independent of the
crossing number and all terms contribute equally giving back shuffle
    \beq
    (F *_0 G)_I = \sum_{I = J \sqcup K} F_J G_K.
    \eeq
{\flushleft \bf Examples:} For $q \ne 1$, $*_q$ is non-commutative
in general. The first few terms in the $q$-product of a pair of
tensors are $(F *_q G)_0 = F_0 G_0$,
    \beq
    (F *_q G)_i &=& F_i G_0 + F_0 G_i, ~~~
        (F *_q G)_{ij} = F_0 G_{ij} + F_i G_j + p F_j G_i + F_{ij}
        G_0, \cr
    (F *_q G)_{ijk} &=& F_0 G_{ijk} + F_i G_{jk} + p F_j G_{ik} + p^2
        F_k G_{ij} + F_{ij} G_k + p F_{ik} G_j + p^2 F_{jk} G_i +
        F_{ijk} G_0 \cr
    (F *_q G)_{ijkl} &=& F_0 G_{ijkl} + (F_i G_{jkl} + p F_j G_{ikl}
    + p^2 F_k G_{ijl} + p^3 F_l G_{ijk} ) \cr &&  + (F_{ij} G_{kl} + p F_{ik} G_{jl}
    + p^2 F_{il} G_{jk} + p^2 F_{jk} G_{il} + p^3 F_{jl} G_{ik}
    + p^4 F_{kl} G_{ij}) \cr && + (F_{ijk} G_l + p F_{ijl} G_k + p^2 F_{ikl} G_j
    + p^3 F_{jkl} G_i) + F_{ijkl} G_0.
    \label{e-eg-q-prod-multi-mat}
    \eeq
{\bf Associativity: }We show that the $q$-product is associative
    \beq
    ((F *_q G) *_q H)_I = (F *_q (G *_q H))_I = \sum_{I = J
        \sqcup K \sqcup L} p^{\chi(I;J,K,L)} F_J G_K H_L.
    \label{e-associativity-of-q-prod}
    \eeq
We first checked explicitly that associativity holds for $|I| \leq
3$ by writing out all the terms, but it was very tedious to go
further. Instead, we write
    \beq
    ((F * G) * H)_I &=& \sum_{I= J \sqcup K} p^{\chi(I;J,K)} (F*G)_J G_K
    = \sum_{I = L \sqcup M \sqcup K} p^{\chi(I; L \sqcup M, K)}
    p^{\chi(L \sqcup M; L,M)} F_L G_M H_K \cr
    &=&  \sum_{I= J \sqcup K \sqcup L}
        p^{\chi(I; J \sqcup K,L) + \chi(J \sqcup K; J,K)} F_J G_K
        H_L  \cr
    (F *(G*H))_I &=& \sum_{I = J \sqcup K \sqcup L} p^{\chi(I;J,K \sqcup L)
        + \chi(K \sqcup L; K,L)} F_J G_K H_L
    \eeq
where $I = J \sqcup K \sqcup L$ is the condition that $J,K,L$ are
complementary order-preserving sub-words of $I$. Since $F,G,H$ are
arbitrary and so is $p$, associativity requires the equality of the
sums of crossing numbers
    \beq
    \chi(I;J \sqcup K, L) + \chi(J \sqcup K; J,K) {\rm ~~and~~}
    \chi(I; J,  K \sqcup L) + \chi(K \sqcup L; K,L)
    \eeq
for each $I$ and any (fixed) choices of $J,K,L, J \sqcup K$ and $K
\sqcup L$ satisfying $I = J \sqcup K \sqcup L$. In fact, these two
sums of (two-word) crossing numbers are equal to the (three-word)
crossing number $\chi(I;J,K,L) $ that has a simple meaning.
$\chi(I;J,K,L)$ is the smallest number of transpositions needed to
transform $JKL$ into $I$ where $J,K,L$ are order-preserving
sub-words of $I$. For example suppose $I = abcd$, $J = d, K= c,
L=ab, J \sqcup K = cd$ and $K \sqcup L = abc$. Then
$\chi(abcd;cd,ab) + \chi(cd;d,c) = 4 + 1 = 5$ while
$\chi(abcd;d,abc) + \chi(abc;c,ab) = 3 + 2 = 5$. Similarly, if $I =
abcd$, $J = b, K=ad, L=c, J \sqcup K =a bd$ and $K \sqcup L = acd$.
Then $\chi(abcd;abd,c) + \chi(abd;b,ad) = 1 + 1 = 2$ while
$\chi(abcd;b,acd) + \chi(acd;ad,c) = 1 + 1 = 2$. Thus, associativity
just says that there are two different ways of calculating the
three-word crossing number $\chi(I;J,K,L)$ when $I = J \sqcup K
\sqcup L$. This gives the simple formula
(\ref{e-associativity-of-q-prod}) for the $*_q$ product of three
series, which makes associativity manifest.

{\flushleft \bf Reduction to one generator:} When we reduce to a
single generator in the above examples
(\ref{e-eg-q-prod-multi-mat}), the formulae agree with those
obtained earlier (\ref{e-eg-q-shuffle-one-mat}) using the Gauss
binomials. More generally, we can see from the definition of the
Gauss binomials (\ref{e-def-gauss-binomials}) that
    \beq
    {|I| \choose r}_q = \sum_{\stackrel{I = J \sqcup K}{|J|=r}}
        q^{\chi(I;J,K)}.
    \eeq
Thus, the above formula for the $q$-product reduces to the one for a
single generator.

{\flushleft \bf Poisson Bracket:} It may help to find a Poisson
bracket on the shuffle algebra that serves as a first approximation
to the $q$-commutator. The $q$-commutator is
    \beq
    ([F,G]_q)_I \equiv (F *_q G - G *_q F)_I = \sum_{I = J \sqcup K}
    (1-q)^{\chi(I;J,K)} (F_J G_K - G_J F_K).
    \eeq
For small $q$, $- \ov{q}([F,G]_q)_I =  \sum_{I = J \sqcup K}
\chi(I;J,K) (F_J G_K - G_J F_K ) + {\cal O}(q).$ So let us define
the bracket $\{F,G\} = \{F,G\}_I \xi^I$ by
    \beq
    \{ F,G\}_I = - \lim_{q \to 0} \ov{q}([F,G]_q)_I = \sum_{I = J \sqcup K} \chi(I;J,K)
        (F_J G_K - G_J F_K ).
    \label{e-def-poisson-bracket}
    \eeq
It is clearly bilinear and anti-symmetric. The first few examples
with lowest $|I|$ are
    \beq
    \{F,G\}_0 &=& 0; ~~~ \{F,G\}_i = 0; ~~~
    \{F,G\}_{ij} = F_j G_i - G_j F_i; \cr
    \{F,G\}_{ijk} &=& F_j G_{ik} + 2 F_k G_{ij} + F_{ik} G_j + 2 F_{jk} G_i
        - (F \leftrightarrow G);  \cr
    \{F,G\}_{ijkl} &=& F_j G_{ikl} + 2 F_k G_{ijl} + 3 F_l G_{ijk} +
    F_{ik} G_{jl} + 2 F_{il} G_{jk} + 2 F_{jk} G_{il} \cr && + 3 F_{jl}
    G_{ik} + 4 F_{kl} G_{ij} + F_{ijl} G_k + 2 F_{ikl} G_j + 3
    F_{jkl} G_i - (F \leftrightarrow  G).
    \eeq
It satisfies the Jacobi identity since the $q$-product was
associative.
    \beq
    \{\{ F,G\}, H\} + \{\{ H,F\}, G\} + \{\{ G,H\}, F\} = 0.
    \eeq
This can also be checked explicitly. For example, the first
non-trivial case is
    \beq
    \{\{ F,G\}, H\}_{ijk} = 2(F_i G_j H_k + F_k G_j H_i
        - F_j G_k H_i - F_j G_i H_k).
    \eeq
Upon adding its cyclic permutations, the Jacobi identity is
satisfied. Moreover, the Leibnitz rule (with respect to $sh =\circ =
*_0$)
    \beq
    \{F \circ G, H \} = F \circ \{G,H\} + \{ F,H\} \circ G
    \eeq
is also satisfied due to the corresponding identity for the
$q$-commutator. Thus $\{...\}$ is a Poisson bracket on the
commutative shuffle algebra.

In order to be practically useful in going beyond the $0^{\rm th}$
order solution of the LE, we need a $q$-expansion for $*_q$ around
$*_0 = sh$ involving left annihilation $D_j^0$. For small $q$,
    \beq
    (F *_q G)_I &=& \sum_{I = J \sqcup K} (1-q)^{\chi(I;J,K)} F_J
        G_K = \sum_{I = J \sqcup K} F_J G_K - q \sum_{I = J \sqcup
        K} \chi(I;J,K) F_J G_K + {\cal O}(q^2) \cr
    &=& (F *_0 G)_I - q \sum_{I = J \sqcup
        K} \chi(I;J,K) F_J G_K + {\cal O}(q^2) \cr
   \Rightarrow && \lim_{q \to 0} {(F *_q G - F *_0 G)_I \over - q}
    = \sum_{I = J \sqcup K} \chi(I; J,K) F_J G_K.
    \eeq
For example,
    \beq
    \lim_{q \to 0} {(F *_q G - F *_0 G)_{ij} \over - q} &=&
    F_j G_i \cr
    \lim_{q \to 0} {(F *_q G - F *_0 G)_{ijk} \over - q} &=&
    F_j G_{ik} + 2 F_k G_{ij} + F_{ik} G_j + 2 F_{jk} G_i.
    \eeq
Our aim is to express this ${\cal O}(q)$ contribution to $F *_q G$
in terms of $D_i^0$ and $*_0$. But we are yet to find such a formula
that generalizes (\ref{e-1-mat-q-series-for-q-prod}) and hope
further investigation will reveal it.

\subsubsection{$q$-Deformed annihilation}
\label{s-q-annihilation}

There is one parameter family of annihilation operators $D_j^q$ that
interpolates between left annihilation $D_j^0$ and full annihilation
$D_j^1$. For a single generator, it was defined in
(\ref{e-1-mat-q-annihilation}) as $(D_q G)_n = (1 + q + q^2 + \cdots
+ q^n ) G_{n+1}$. By analogy we define $[D_j^q G]_I = \delta_I^{I_1
I_2} q^{|I_1|} G_{I_1 j I_2}$, i.e.
    \beq
    [D_j^q G]_{i_1 \cdots i_n} = G_{j i_1 \cdots i_n} +
        q G_{i_1 j i_2 \cdots i_n} + q^2 G_{i_1 i_2 j i_3 \cdots
        i_n} + \cdots + q^n G_{i_1 \cdots i_n j}.
    \eeq
We pick up one more power of $q$ as the annihilation operator
travels through each index of the tensor from left to right. It is
easily seen that
    \beq
    \lim_{q \to 0} ~[D_j^q G]_I = G_{jI} {~~ \rm and ~~}
    \lim_{q \to 1} ~[D_j^q G]_I = \delta_I^{I_1 I_2} G_{I_i j I_2}
    \eeq
reproduce left and full annihilation which are derivations of $sh$
and $conc$. To make the LE (\ref{e-loop-eqns-in-terms-of-left-ann})
differential equations with respect to $conc$, we want to expand
$D^0_j$ around $D^1_j$ in powers of $p=1-q$ and finally set $p=1$.
Recall that for $1$-generator
(\ref{e-1-mat-q-annihilation-expanded-in-p}),
    \beq
    D_q G(\xi) &=& \sum_{k=1}^\infty \ov{k!} (-p\xi)^{k-1} D_0^k G(\xi)
        = D_1 G(\xi) - {p \over 2} \xi D_1^2 G(\xi)
        + {p^2 \over 6} \xi^2 D_1^3 G(\xi) + {\cal O}(p^3).
    \eeq
For several generators,
    \beq
    [D_j^q G]_{i_1 \cdots i_n}
        &=& \bigg[G_{j i_1  \cdots i_n} + \cdots + G_{i_1 \cdots i_n j}
            \bigg]
        - p \bigg[G_{i_1 j i_2 \cdots i_n} + 2 G_{i_1 i_2 j i_3 \cdots
            i_n} + \cdots + n G_{i_1 \cdots i_n j} \bigg] \cr
        + p^2 \bigg[G_{i_1 i_2 j i_3 \cdots i_n} &+& 3 G_{i_1 i_2 i_3 j i_4 \cdots i_n}
        + \cdots + {n(n-1) \over 2} G_{i_1 \cdots i_n j} \bigg]
        + \cdots + (-p)^n G_{i_1 \cdots i_n j}.
    \eeq
Drawing inspiration from
(\ref{e-1-mat-q-annihilation-expanded-in-p}) we would like to
recognize the coefficients of powers of $p$ as combinations of full
annihilation and some multiplication operator acting on $G$.
However, we have not yet succeeded in this.


\section{Discussion}
\label{s-discussion}

Despite their formidable reputation, the loop equations(LE) of a
large-$N$ multi-matrix model show much simplicity and structure when
expressed in terms of gluon correlations $G_I$. Non-linearities are
mild in the sense that in any equation, highest rank correlations
appear linearly. So the LE are systems of inhomogeneous linear
difference equations for correlations of a given rank with lower
rank correlations appearing non-linearly as `sources'. Solving these
equations in the absence of additional structure would be tedious at
best. But this is not possible because the LE are underdetermined in
most interesting cases. We observed that there are additional
equations involving the $G_I$ that a naive passage from finite $N$
Schwinger-Dyson equations to large-$N$ LE misses. These equations
have to do with changes of variables in matrix integrals that leave
both action and measure invariant. However, we are yet to implement
these additional constraints in detail to see whether they suffice
to fix a unique solution to the LE. On the other hand, we saw that
part of the difficulty in understanding the LE lies in the fact that
they are not differential equations. Left annihilation does not
satisfy the Leibnitz rule with respect to the concatenation product
appearing in these equations. We proposed two schemes to remedy this
situation by expanding either annihilation or product around one
that is a derivation of the other. For the Gaussian, Chern-Simons
and Yang-Mills models, it was possible to altogether eliminate the
non-linearities of the LE and arrive at inhomogeneous linear PDEs at
the zeroth order of these expansions. But the under-determinacy of
the loop equations prevented us from picking a unique solution
except in the case of the gaussian, where the two approximations
were shown to give over and underestimates for correlations. This
underscores the importance of better understanding the remaining
constraints on $G_I$ (section \ref{s-additional-eqns}) as well as
any other conditions that would ameliorate the under-determinacy of
the LE. In \cite{gsk-ghost-shuffle-concat} we hope to extend these
algebraic and differential properties to matrix models with both
gluon and ghost matrices, of the sort appearing in the gauge-fixed
action of Yang-Mills theory.

\section*{\normalsize Acknowledgements}

The author has benefitted from numerous discussions with S. G.
Rajeev, A. Agarwal and L. Akant, for which he is very grateful. The
author also thanks G. Arutyunov, A. Cattaneo and G. Felder for
discussions and acknowledges support of the European Union in the
form of a Marie Curie Fellowship. Thanks are also due to G. 't Hooft
for encouragement to `devise more powerful calculation techniques'
for Yang-Mills theory.

\appendix

\section{Cyclically symmetric tensors of rank $n$}
\label{a-cyclic-tensors}

What is the dimension $c(n,\La)$ of the space of cyclically
symmetric real tensors $G_{i_1 \cdots i_n}$ of rank (=number of
indices) $n$ if the indices can take the values $1 \leq i_k \leq
\La$? The dimension of the space of all tensors of rank $n$ is
$\La^n$. On the other hand, the space of symmetric rank $n$ tensors,
which is a subspace of cyclically symmetric tensors, is ${\La + n -1
\choose n}$ dimensional. Thus
    \beq
    {\La + n -1 \choose n} ~\leq~ c(n,\La) ~\leq~ \La^n
    \eeq
For a $\La = 3$ matrix model, $\half(n^2 + 3n +2) \leq~ c(n,3) \leq
3^n$. For a $2$-matrix model, $n+1 \leq c(n,2) \leq 2^n.$ The cyclic
group of order $n$ acts on rank $n$ tensors $G_{i_1 \cdots i_n}$ by
cyclically permuting indices. $c(n,\La)$ is the number of orbits.
For example, if $\La = 2$ and $n=4$, the orbits are
    \beq
    (G_{2222}); && (G_{1222} = G_{2122} = G_{2212} = G_{2221});~~~
    (G_{1122}=G_{2112}=G_{2211}=G_{1221}); ~~~ \cr
    (G_{1212}=G_{2121}); &&
    (G_{1112}=G_{2111}=G_{1211}=G_{1121});~~~ (G_{1111})
    \eeq
So $c(n=4,\La=2) = 6$, significantly less than $2^4$. The
cardinality of different orbits are not necessarily equal. Some
other examples are
    \beq
    c(n,1) = 1; &&
    c(1,\La) = \La; ~~~
    c(2,\La) = \half \La (\La +1); ~~~
    c(3,2) = 4; \cr
    c(3,3) = 11; &&
    c(4,2) = 6; ~~~
    c(4,3) = 24;~~~
    c(5,2) = 8.
    \eeq
It would be nice to have formula for $c(n,\La)$, at least for the
$\La = 2$ matrix model.

{\flushleft \bf Note on hermiticity condition:} Actually, the
tensors $G_I$ are complex numbers, so the real-dimension of the
space of cyclically symmetric tensors of rank $n$ is $2~ c(n,\La)$.
However, the hermiticity condition $G_{i_1 i_2 \cdots i_n} =
G^*_{i_n \cdots i_2 i_1}$ halves this real-dimension to $c(n,\La)$.
If reversal of indices can be achieved by a cyclic permutation (e.g.
$G_{1122}=G_{2211}^* = G_{2211}$) then the correlation is real. If
$\bar I$ cannot be obtained from $I$ via cyclic permutations, then
hermiticity means that $\Re G_I = \Re G_{\bar I}$ and $\Im G_I = -
\Im G_{\bar I}$. For example $\Re G_{1123} = \Re G_{3211}$ and $\Im
G_{1123} = -\Im G_{3211}$. In either case, hermiticity halves the
number of independent parameters in cyclically symmetric
correlations of a given rank.

\section{Concatenation, shuffle and their co-products}
\label{a-conc-sh-and-coprods}

By $V$, let us denote the infinite-dimensional complex vector space
spanned by the monomial words $\xi^{i_1 \cdots i_n}$ in the $\La$
non-commuting sources $\xi^i$. A typical element is the formal
series $G(\xi) = G_I \xi^I$. $V$ is the basic arena for our
algebraic study of the loop equations\footnote{A superior approach
that makes cyclic symmetry of $G_I$ manifest might be to consider
the quotient by the relation $\xi^I \sim \xi^J$ if $I$ is a cyclic
permutation of $J$. Then a basis for $V$ would consist of words
$\xi^I$ where $I$ labels orbits of the cyclic group action. In this
paper we just allow all words $\xi^I$ and impose the condition that
$G_I$ be cyclically symmetric, by hand, so to speak.}.

The concatenation product $conc : V \otimes V \to V$ denoted by
juxtaposition, was defined in (\ref{e-concat-prod}) $\xi^I \xi^J =
\delta^{IJ}_K \xi^K = \xi^{IJ}$. It has the structure constants
$c^{I,J}_K = \delta^{IJ}_K$. For $\La > 1$, $conc$ is
non-commutative. The vector space $V$, along with the concatenation
product is the free associative algebra ${\cal T}$ on the generators
$\xi^{1}, \cdots, \xi^{\La}$. It is the universal envelope of the
free Lie algebra. The commutative shuffle product $sh: V \otimes V
\to V$ was defined in (\ref{e-sh-prod-of-gluon-corrlns}). $V$,
equipped with $sh$ is the shuffle algebra. The shuffle product of
monomials
    \beq
    \xi^I \circ \xi^J = s^{I,J}_K \xi^K = \sum_{I \sqcup J = K}
    \xi^K.
    \eeq
leads to the shuffle structure constants $s^{I,J}_K = |\{I \sqcup J
= K\}|$.

There is a natural inner product $( .,. )$ on $V$, for which $\xi^I$
form an orthonormal basis
    \beq
    ( \xi^I, \xi^J ) = \delta^{I,J}  {\rm ~~~or~~~} ( F_I \xi^I,
    G_J \xi^J ) = F_I G_J \delta^{I,J} = \sum_{I} F_I G_I.
    \eeq
The Kronecker symbol $\d^{I,J} = 1$ if $I = J$ and $0$ otherwise. We
can use the `metric' $\d^{I,J}$ and its inverse $\d_{I,J}$ to raise
and lower indices. The inner product allows us to define co-products
$V \to V \otimes V$. We call them co-concatenation $\D = sh^\dag$
and co-shuffle $\D' = conc^\dag$. They are adjoints of $sh$ and
$conc$ respectively. For three formal series $F,G,H$, we define $\D$
and $\D'$ by
    \beq
    ( F \otimes G, \D(H) ) = ( F \circ G , H )
    ~~~ {\rm and}  ~~~
    ( F \otimes G, \D'(H) ) =( FG , H ).
    \eeq
We define the structure constants of co-concatenation and co-shuffle
as
    \beq
    \D(\xi^K) = s^K_{L,M} \xi^L \otimes \xi^M
    {~~\rm and~~} \D'(\xi^K) = c^K_{L,M} \xi^L \otimes \xi^M.
    \eeq
We use the same letter $c$ to denote the structure constants of
$conc$ and $\D' = conc^\dag$ because they are related by raising and
lowering indices using the metric $\d^{I,J}$. The same goes for the
letter $s$ for the structure constants of $sh$ and $\D = sh^\dag$.
The expressions for these are
    \beq
    c^I_{J,K} = c^{L,M}_N \d^{I,N} \d_{J,L} \delta_{K,M} = \delta^I_{JK} {~~\rm and~~}
    s^I_{J,K} = s^{L,M}_N \d^{I,N} \d_{J,L} \delta_{K,M} = s^{J,K}_I
        = |\{I= J \sqcup K\}|.
    \eeq
To obtain the co-shuffle structure constants $c^I_{J,K}$ , we use
the definition of adjoint to get
    \beq
    \bra \xi^I \xi^J , \xi^K \ket &=&  \bra \xi^I \otimes \xi^J ,
        \D'(\xi^K) \ket ~~
    \Rightarrow~~ \d^{IJ,K} = c^K_{L,M} \bra \xi^I \otimes \xi^J ,
        \xi^L \otimes \xi^M \ket =  c^K_{L,M} \d^{I,L} \d^{J,M} \cr
    \Rightarrow~~ c^K_{N,P} &=& \d^{IJ,K} \d_{I,N} \d_{J,P} =
    \d^K_{NP}.
    \eeq
We use a similar procedure for the co-concatenation structure
constants $s^I_{J,K}$
    \beq
    \bra \xi^I \circ \xi^J , \xi^K \ket &=&  \bra \xi^I \otimes \xi^J ,
        \D(\xi^K) \ket ~~~
    \Rightarrow~~ s^{I,J}_{L} \bra \xi^L, \xi^K \ket  = s^K_{L,M} \bra \xi^I \otimes \xi^J ,
        \xi^L \otimes \xi^M \ket \cr
    \Rightarrow ~~~ s^{I,J}_{L} \d^{L,K} &=& s^K_{L,M} \d^{I,L}
    \d^{J,M} ~~ \Rightarrow ~~ s^K_{P,Q} = s^{I,J}_L \d^{L,K}
    \d_{I,P} \d_{J,Q} = s^{P,Q}_K.
    \eeq
On formal series, co-shuffle $\D' = conc^\dag$ acts as
    \beq
    \D' F = [\D' F]_{I,J} ~\xi^I \otimes \xi^J
        = F_{IJ} \xi^I \otimes \xi^J.
    \eeq
In particular, $\D'(\xi^I) = \delta^I_{JK} \xi^J \otimes \xi^K$ and
$\D'(\xi^i) = (\xi^i \otimes  1 + 1 \otimes \xi^i)$. On formal
series, co-concatenation $\D = sh^\dag$ acts according to
    \beq
    \D F = [\D F]_{J,K} \xi^J \otimes \xi^K {\rm ~~~where~~~}
    [\D F]_{J,K}  = \sum_{I = J \sqcup K} F_I.
    \eeq
In particular, $\D(\xi^I) = \sum_{I = J \sqcup K} \xi^J \otimes
\xi^K$ and $\D(\xi^i) = \xi^i \otimes 1 + 1 \otimes \xi^i.$

\section{Bialgebra structures on $V = Span(\xi^I)$}
\label{a-bialgebra-structure}

$V$ has two bialgebra (algebra $+$ compatible coalgebra) structures.
In one, the product ($sh$) is commutative while the co-product
(adjoint of $conc$) is non-co-commutative. In the dual bialgebra,
the product ($conc$) is non-commutative while the co-product
(adjoint of $sh$) is co-commutative.

To establish that shuffle and co-shuffle\footnote{This justifies the
name co-shuffle for the adjoint of $conc$.} combine to define a
bialgebra on $V$, we show that co-shuffle $\D' = conc^\dag$ is a
homomorphism of the shuffle product
    \beq
    \D'(F \circ G) = \D'(F) \circ \D'(G).
    \eeq
Note that the LHS is
    \beq
    \D'(F \circ G) = \sum_{L \sqcup M = K} F_L G_M \D'(\xi^K)
        = \sum_{L \sqcup M = IJ} F_L G_M \xi^I \otimes
            \xi^J.
    \eeq
While the RHS is
    \beq
    \D'(F) \circ \D'(G) &=& F_I G_J \D'(\xi^I) \circ \D'(\xi^J)
    = F_I G_J \d^I_{KL} \d^J_{MN} (\xi^K \otimes \xi^L) \circ
        (\xi^M \otimes \xi^N) \cr
    &=& F_{KL} G_{MN} (\xi^K \circ \xi^M) \otimes (\xi^L \circ \xi^N)
    = \sum_{K \sqcup M = I, L \sqcup N = J} F_{KL} G_{MN} \xi^I
        \otimes \xi^J.
    \eeq
Comparing coefficients, $\D' = conc^\dag$ is a homomorphism of the
shuffle product if
    \beq
    \sum_{J_1 \sqcup J_2 = I_1 I_2} F_{J_1} G_{J_2}
        = \sum_{L_1 \sqcup M_1 = I_1, L_2 \sqcup M_2
        = I_2} F_{L_1 L_2} G_{M_1 M_2} ~~~~ \forall~~ I_1, ~I_2.
    \eeq
To prove this, observe that $J_1$ may be uniquely decomposed as $J_1
= L_1 L_2$ with $L_1 \subset I_1$ and $L_2 \subset I_2$ and
similarly for $J_2$, $J_2 = M_1 M_2$ with $M_1 \subset I_1$ and $M_2
\subset I_2$. Then we observe that every riffle-shuffle $J_1 \sqcup
J_2 = I_1 I_2$ arises from a unique pair of riffle-shuffles $L_1
\sqcup M_1 = I_1$ and $L_2 \sqcup M_2 = I_2$. This establishes that
co-shuffle $\D$ is a homomorphism of $sh$.

A similar argument shows that $\D = sh^\dag$ is a homomorphism of
$conc$: $\D(FG)= \D(F) \D(G)$.
    \beq
    \D(F) \D(G) = \sum_{J = I_1 I_3, K = I_2 I_4} (\D F)_{I_1, I_2} (\D
    G)_{I_3, I_4} = \sum_{J = I_1 I_3, K= I_2 I_4} F_{I_1 \sqcup I_2}
    G_{I_3 \sqcup I_4}.
    \eeq
On the other hand, the LHS gives
    \beq
    [\D(FG)]_{J,K} = \sum_{L = J \sqcup K} (FG)_L = \sum_{L_1 L_2 = J
        \sqcup K} F_{L_1} G_{L_2}
    = \sum_{J = I_1 I_3, K= I_2 I_4} F_{I_1 \sqcup I_2} G_{I_3 \sqcup
    I_4}.
    \eeq
In the last equality, we used the unique decomposition $J = I_1 I_3,
K = I_2 I_4$ where $I_1, I_2 \subset L_1$ and $I_3, I_4 \subset L_2$
as before. Thus we have shown that $\D = sh^\dag$ is a homomorphism
of $conc$.

The unit element for $conc$ is $1$, $F 1 = 1 F = F$. The co-unit for
co-concatenation is $\eps : V \to \mathbf{C}$. It picks out the
constant term in a formal series $\eps(F_I \xi^I) = F_\emptyset
\equiv F_0$. Just like co-concatenation, the co-unit is a
homomorphism of $conc$
    \beq
    \eps(FG) = (FG)_0 = F_0 G_0 = \eps(F) \eps(G).
    \eeq
The unit element for shuffle too is $1$, $(F \circ 1)_I = \sum_{I =
J \sqcup K} F_J \delta_K^0 = F_I$. The co-unit for co-shuffle is
again $\eps: V \to \mathbf{C}$. The co-unit $\eps$ is a homomorphism
of the shuffle product
    \beq
    \eps(F \circ G) = (F \circ G)_0 = \sum_{J \sqcup K = \emptyset}
    F_J G_K = F_0 G_0 = \eps(F) \circ \eps(G).
    \eeq
To summarize, $(conc, sh^\dag = \D = co{\rm-}conc, 1, \eps)$ defines
a non-commutative but co-commutative bialgebra (algebra plus
compatible co-algebra) structure on $V=span(\xi^I)$. Similarly,
$(sh, conc^\dag = \D' = co{\rm-}sh, 1, \eps)$ defines a commutative
but non-co-commutative bialgebra structure on $V$. These two
bialgebras are not independent. Structure constants of the product
and co-product of one can be obtained from those of the other using
the inner product $\delta^{I,J}$ on $V$.

{\flushleft \bf Remark:} In addition to being a bialgebra ${\cal T}
= (conc, \D,1,\eps)$, is the universal envelope of the free Lie
algebra. So it is a {\em Lie} algebra with the Lie product
$[\xi^I,\xi^J] = \xi^{IJ} - \xi^{JI}$. Does $\D$ define a {\em Lie}
bialgebra \cite{chari-pressley} with respect to the commutator? No!
On the one hand, $\D: V \to V \otimes V$ is not skew-symmetric.
Rather, its image lies within $Sym(V \otimes V)$.
    \beq
    \D(\xi^I) = \sum_{I = J \sqcup K} \xi^J \otimes \xi^K =
    \sum_{I = J \sqcup K} \xi^K \otimes \xi^J = (\tau \D)(\xi^I),
    \eeq
where $\tau(a \otimes b) = b \otimes a$. Here we used the fact that
if $J$ and $K$ are order preserving complementary subwords of $I$,
then so are $K$ and $J$. Furthermore, $\D$ is not a 1-cocycle for
the free associative algebra. In order to be a 1-cocycle, it must
satisfy
    \beq
    \D[F,G] = (ad_F \otimes 1 + 1 \otimes ad_F) \D(G) -
    F \leftrightarrow G
    \eeq
for any $F,G \in {\cal T}$. However, taking $F = \xi^i$ and $G =
\xi^j$ gives
    \beq
    LHS = \D[\xi^i,\xi^j] = \xi^{ij} \otimes 1 + 1 \otimes \xi^{ij} -
    \xi^{ji} \otimes 1 - 1 \otimes \xi^{ji}
    \eeq
and $RHS = 2 \times LHS \ne LHS$. There may be some other
skew-symmetric 1-cocycle $\tl\D: {\cal T} \to {\cal T} \otimes {\cal
T}$ which defines a Lie bialgebra structure on the universal
envelope of the free Lie algebra.

\footnotesize



\end{document}